\documentclass[preprint]{aastex}
\usepackage{natbib,ifthen}
\usepackage{graphicx}
\usepackage{lscape}
\slugcomment{}
\newcommand {\mic} {$\mu$m}
\newcommand {\ben} {\begin{eqnarray}}
\newcommand {\een} {\end{eqnarray}}
\newcommand {\msun} {M$_\odot$}
\newcommand{\msunperyr}{M$_\odot$ yr$^{-1}$}
\newcommand{\kss}{K$_{\rm s}$}
\newcommand{\ks}{K$_{\rm s}$ }
\newcommand{\ltapp}{\raisebox{-.4ex}{\rlap{$\sim$}} \raisebox{.4 ex}{$<$}}

\shortauthors{Srinivasan {\it et al.}}
\shorttitle{SAGE Evolved Stars Infrared Excesses}
\begin{document}
   \title{The Mass-Loss Return From Evolved Stars to the LMC:
   Empirical Relations for Excess Emission at 8 and 24 $\mu$m.}
   
   \author{Sundar Srinivasan\altaffilmark{1}, Margaret Meixner\altaffilmark{2}, Claus Leitherer\altaffilmark{2}, Uma Vijh\altaffilmark{2}, 
   Kevin Volk\altaffilmark{3}, Robert D. Blum\altaffilmark{4}, Brian L. Babler\altaffilmark{5}, Miwa Block\altaffilmark{6}, Steve Bracker\altaffilmark{5}, Martin Cohen\altaffilmark{7}, Charles W. Engelbracht\altaffilmark{6}, Bi-Qing For\altaffilmark{6},  Karl D. Gordon\altaffilmark{2}, Jason Harris\altaffilmark{4}, Joseph L. Hora\altaffilmark{8}, Remy Indebetouw\altaffilmark{9}, Francisca Markwick--Kemper\altaffilmark{10}, Marilyn Meade\altaffilmark{5}, Karl A. Misselt\altaffilmark{6}, Marta Sewilo\altaffilmark{2}, Barbara Whitney\altaffilmark{11}}

  \altaffiltext{1}{Department of Physics and Astronomy,
              The Johns Hopkins University,
              3400 North Charles St., Baltimore, MD 21218, USA, email: {\tt sundar@pha.jhu.edu}
              }
\altaffiltext{2}{Space Telescope Science Institute,
	      3700 San Martin Drive,
	      Baltimore, MD 21218, USA
	       }
\altaffiltext{3}{Gemini Observatory,
	      670 North A'ohoku Place,
	      Hilo, HI 96720, USA
	      }
\altaffiltext{4}{National Optical Astronomy Observatory, PO Box
	      26732, Tucson AZ 85726-6732, USA
	   }
\altaffiltext{5}{Department of Astronomy, University of Wisconsin, 475 North Charter Street, Madison, WI 53706, USA
	   }	   
\altaffiltext{6}{Steward Observatory, University of Arizona, 933 North Cherry Avenue, Tucson, AZ 85721, USA
	   }
\altaffiltext{7}{Radio Astronomy Laboratory, 601 Campbell Hall, University of California at Berkeley, Berkeley, CA 94720, USA	   }
\altaffiltext{8}{Harvard-Smithsonian Center for Astrophysics, 60 Garden Street, Cambridge, MA 02138, USA
	   }
\altaffiltext{9}{Astronomy Department, University of Virginia, P.O. Box 3818, Charlottesville, VA 22903-0818, USA
	   }
\altaffiltext{10}{Jodrell Bank Centre of Astrophysics, School of Physics \& Astronomy, University of Manchester, Manchester, M13 9PL, UK	  
	     }	   	   
\altaffiltext{11}{Space Science Institute, 4750 Walnut Street, Suite 205, Boulder, CO 80301, USA
	   }

\begin{abstract}
We present empirical relations describing excess emission from evolved stars 
in the Large Magellanic Cloud (LMC) using data from the Spitzer Space Telescope SAGE 
(Surveying the Agents of a Galaxy's Evolution) survey which includes the IRAC 3.6, 4.5, 5.8 and 8.0 \mic\ and MIPS 24, 70 and 160 \mic\ bands. We combine the SAGE data with the Two Micron All Sky Survey (2MASS; J, H and \kss)
and the optical Magellanic Cloud Photometric Survey (MCPS; U, B, V and I) point source catalogs in order to create complete spectral energy distributions (SEDs) of the asymptotic giant branch (AGB) star candidates in the LMC. AGB star outflows are among the main producers of dust in a galaxy, and this mass loss results in an excess in the fluxes observed in the 8 and 24 \mic\ bands. The aim of this work is to investigate the mass-loss return by AGB stars to the interstellar medium of the LMC by studying the dependence of the infrared excess flux on the total luminosity. We identify oxygen-rich, carbon-rich and extreme AGB star populations in our sample based on their 2MASS and IRAC colors. The SEDs of oxygen- and carbon-rich AGB stars are compared with appropriate stellar photosphere models to obtain the excess flux in all the IRAC bands and the MIPS 24 \mic\ band. Extreme AGB stars are dominated by circumstellar emission at 8 and 24 \mic\, thus we approximate their excesses with the flux observed in these bands. We find about 16,000 O--rich, 6300 C--rich and 1000 extreme sources with reliable 8 \mic\ excesses, and about 4500 O--rich, 5300 C--rich and 960 extreme sources with reliable 24 \mic\ excesses. The excesses are in the range 0.1 mJy -- 5 Jy. The 8 and 24 \mic\ excesses for all three types of AGB candidates show a general increasing trend with luminosity. The color temperature of the circumstellar dust derived from the ratio of the 8 and 24 \mic\ excesses decreases with an increase in excess, while the 24 \mic\ optical depth increases with excess. The extreme AGB candidates are the major contributors to the mass loss, and we estimate the total AGB mass-loss return to the LMC to be $(5.9-13)\times 10^{-3}$ \msunperyr.
\end{abstract}


\section{Introduction}
\noindent
Towards the final stages of their evolution, low- and intermediate-mass stars (0.8 -- 8 \msun)
enter the asymptotic giant branch (AGB) phase. Characterized by cool ($\sim$3000 K) photospheric temperatures and dusty circumstellar (CS) shells which absorb optical light from the star and re-emit it at longer wavelengths, AGB stars are among the brightest infrared (IR) sources in a galaxy. During the AGB phase, a significant fraction of the star's original mass is lost at rates of up to $\sim$10$^{-4}$ \msunperyr\ \citep[][]{WoodBessellFox83, Woodetal92}, resulting in the chemical enrichment of the interstellar medium (ISM).

AGB mass loss is believed to be a two-step process: pulsations first levitate material above the photosphere, where the cool temperatures result in the formation of dust grains. Radiation pressure then drives the dust grains (which are collisionally coupled with the gas) outward in an efficient stellar wind \citep[][]{Wick66,GS76,BowenWillson91,Wachteretal02}. An increase in luminosity (and hence radiation pressure) must therefore, in general, be accompanied by an increased mass-loss rate (MLR). This inference is supported by various observations and model predictions 
\citep[see, {\it e.g.},][]{Reimers75,VW93,Blocker95,BowenWillson91,vL1999,Wachteretal02,vL2005}. 

The formation of a circumstellar envelope (CSE) is a direct result of the mass-loss process. The flux from the CS dust shell appears as mid-infrared (MIR) emission in excess of that due to the central star alone. This MIR excess is directly related to the rate of mass loss, and is therefore expected to increase with increasing luminosity of the central star. Various studies in the past have demonstrated the relationship between the MIR excess and the rate of mass loss. \citet{SW88} showed that the MLR derived from CO rotational transitions was proportional to the strength of the 9.5 $\mu$m silicate dust feature in optically thin O--rich stars. \citet{Jura87} derived a relation between the MLR and 60 $\mu$m excess for O-- and C--rich AGB stars in the solar neighborhood. \citet{KGW92} found that the 12 $\mu$m emission from elliptical galaxies was proportional to the measured MLR. \citet{Athey02} compared the MIR excess emission from 9 galaxies to that of Galactic and LMC AGB stars and derived a proportionality relation between the MIR excess and MLR.

While the study of Galactic AGB stars is hindered by obscuring Galactic dust along the line of sight and high uncertainty in distance estimates, the Large Magellanic Cloud (LMC) is ideal for such observations owing to its proximity \citep[$\sim$50 kpc,][]{Feast99} and favorable viewing angle \citep[35$^\circ$,][]{vdM01}. 
Early LMC surveys \citep[][]{West78,West81,Rebeirot83,BBM80,BM83,FB90} looked for AGB stars  at optical and near-infrared (NIR) wavelengths, and hence preferentially detected sources with optically thin CS dust shells. The Infrared Astronomy Satellite \citep[IRAS;][]{IRAS} survey of the LMC \citep{IRASPSC} in conjunction with ground-based NIR confirmation \citep{Reid90} helped identify several of the brightest mass-losing evolved stars \citep{Loup97,Zijlstra96}. \citet{Trams99} performed follow-up mid-infrared (MIR) photometry and spectroscopy of 57 sources using the  Infrared Space Observatory \citep[ISO;][]{ISO} in order to determine the chemical composition of the CSEs, and 
\citet{vL1999} used radiative transfer modeling to derive luminosities and MLRs for these sources. More recently, the LMC has been surveyed in the optical by the Magellanic Clouds Photometric Survey \citep[MCPS;][]{MCPS}, and in the NIR by DENIS \citep{DENIS} and 2MASS \citep{2MASS}. The LMC data from these two surveys can be found in \citet{Cioni} and \citet{2MASSLMC}, respectively. A mid-infrared survey using the Midcourse Space Experiment \citep[MSX;][]{MSX}, which was four times more sensitive than IRAS, was performed by \citet{MSXLMC}.

As part of the SAGE survey \citep[{\it Surveying the Agents of a Galaxy's Evolution};][]{sage1}, a $\sim$7$^\circ\times 7^\circ$ area of the LMC was imaged in the Spitzer Space Telescope \citep[{\it Spitzer};][]{Spitzer} IRAC (3.6, 4.5, 5.8 and 8.0 $\mu$m) and MIPS (24, 70 and 160 $\mu$m) bands. One of the main goals of SAGE was to detect all the evolved stars with MLRs $> 10^{-8}$ \msunperyr, to characterize the total rate at which material is returned to the ISM by dusty evolved stars, and to better understand the physics governing mass loss among evolved stars in the LMC. When complete, the SAGE data will be $\sim$1000 times more sensitive than MSX, and they will allow a detailed quantitative derivation of the global mass-loss budget from all stellar populations when combined with existing and future MIR spectroscopic observations of evolved stars \citep[see, {\it e.g.},][]{vL1999,vL2005,Zijlstra06}. Continuing the analysis begun by \cite{sage1}, \citet{sage2} identified about 32,000 color-selected evolved stars brighter than the tip of the red giant branch (11.85 mag in the IRAC 3.6 $\mu$m band), including 17,500 oxygen--rich (O--rich), 7000 carbon--rich (C--rich), and 1200 ``extreme" AGB stars, and presented color-magnitude diagrams (CMDs) of the SAGE epoch 1 data.

Our aim is to investigate the MLRs of the AGB candidates selected from the SAGE survey, which will require detailed radiative transfer modeling of the CSE around each such star. In order to simplify this effort to model $\sim 10^4$ stars, we would like to constrain the range of some of the input model parameters, such as the total source luminosity, the typical temperature of the CS dust and the optical depth at a given wavelength. As a first step towards this modeling goal, therefore, in this paper we take an empirical approach and calculate MIR excess fluxes due to dust from these stars. The distance independence of the MIR colors of AGB sources makes them a good choice for tracing the mass loss \citep[see, {\it e.g.},][]{Whitelocketal2006}, but our study based on the MIR excesses for sources in the LMC is aided by the fact that the distance to the stars in the LMC is essentially the same. Data from the SAGE survey also provides an unprecedented opportunity to include a large number of AGB star candidates. We will show that the MIR excess can be used as a proxy for the MLR. In a very statistical way, we study the overall trends of excess and derived quantities such as color temperature of the dust and MIR optical depth with source luminosity. In subsequent papers, we will also be presenting the luminosity function and detailed radiative transfer calculations for the AGB star candidates selected in this paper. The paper is organized as follows: in \S \ref{data} we describe the SAGE database and our observational sample of AGB stars. Our procedure for calculating IR excesses is explained in \S \ref{excess}. We present our results in \S \ref{results}, and in \S \ref{discuss} we compare our results to previous work and discuss their implications for future AGB studies in the LMC.

\section{Data}
\label{data}
\noindent The SAGE catalog and archive point source lists from both epochs of observations have been delivered to the Spitzer Science Center (SSC), and are available for download at the SSC\footnote{http://ssc.spitzer.caltech.edu/legacy/sagehistory.html}. The SAGE epoch 1 point source catalog is discussed by \citet{sage1} and \citet{sage2}. In this study, we select sources from the IRAC epoch 1 archive instead of the catalog. The archive accepts fainter sources than the catalog. Faint limits for both epochs are 18.5, 17.5, 15, and 14.5 mag for IRAC 3.6, 4.5, 5.8, and 8.0 \mic\, respectively, compared to 18, 17, 15, and 14 for the catalog. A source is excluded ({\it i.e.}, culled) from the archive if it has neighbors within a 0.5\arcsec~radius, whereas this radius is 2\arcsec~for the catalog. This culling procedure ensures the creation of reliable lists of point sources at the expense of completeness \citep[see][]{sage1}. In order for two sources to appear blended, they both have to be detected, and they have to be bright enough. The probability of such blending in the 24 \mic\ band is very low, since the 24 \mic\ catalog has very few sources  ($\sim 40,000$) compared to the IRAC catalog. However, we do find a few ($\sim$40 out of $\sim 10^4$) IRAC sources that are matched to more than one MIPS 24 \mic\ source which we corrected in our results and in the main catalog. In addition to this culling, a flux value for each archive source in any of the IRAC bands is non-null only if its signal-to-noise ratio (S/N) is greater than 5. For the catalog, the S/N is 6 for the IRAC [3.6], [4.5], and [5.8] \mic\ bands, and 10 for the [8.0] \mic\ band. As a result of these criteria, the archive has slightly more sources and more flux values than the catalog, allowing the inclusion of more faint AGB candidates. As of the second delivery, version S13, the archive has $\sim$4.5 million sources compared to $\sim$4.3 million in the catalog. See the SAGE data delivery document\footnote{http://data.spitzer.caltech.edu/popular/sage/20080204\_enhanced/documents/SAGEDataDescription\_Delivery2.pdf} for details of the source selection in the catalog and archive.  The fact that the fluxes of these fainter AGB candidates are more uncertain when using the archive list is mitigated by our requirement of 2MASS detections thus affirming that the point source is real. The AGB source lists are extracted from a universal table which lists the nearest neighbor in the MCPS and IRAC epoch 1 archive to each source in the MIPS24 epoch 1 full catalog ($\sim$40,000 sources). The matching radius is 3\arcsec, but the nearest neighbor is used in the match. The IRAC archive sources are also bandmerged with the 2MASS catalog.\\

We classify sources as low- or moderately-obscured O--rich/C--rich AGB candidates based on their location in the \ks versus J-\ks CMD\footnote{The UBVI and JH\ks magnitudes are dereddened using the same procedure as in \citet{Cioni}} \citep{Cioni,sage2}, but we exclude stars without H-band fluxes from our list, as our procedure for calculating the excesses (see \S \ref{excess}) relies on an H-band detection. This criterion probably excludes the faintest O--rich AGB stars, and thus does not significantly affect our results. As a separation into O--rich and C--rich chemistries based on near-IR colors is not possible for the heavily obscured (J--[3.6]$>$3.1, \citet{2MASSLMC}) ``extreme" AGB candidates, we follow a procedure similar to \citet{sage2} and select these sources based on their location in the [3.6] versus J--[3.6] CMD or (when no 2MASS counterpart exists) the [8.0] versus [3.6]--[8.0] CMD. A simple trapezoidal integration of the optical U through MIPS24 fluxes is performed to estimate the luminosities for all our sources.

The \citet{Kastneretal2008} list of objects with spectroscopic classifications contains 14 sources classified in our study as O--rich AGB stars. \citet{Kastneretal2008} classify 10 of these as red supergiants (RSGs). We also find that we have classified 21 of their sources as extreme AGB stars -- most of these have been identified as possible HII regions, none of them are RSGs. Since these are point sources, we interpret HII regions as compact HII regions or massive young stellar objects (YSOs). We thus realize that our sample of AGB stars may be contaminated by non-AGB objects such as RSGs and YSOs. We estimate this contamination using simple color-magnitude cuts. 

The figures in \citet{sage2} also show that the AGB stars and RSGs are not well separated in the MIR CMDs. It is possible that some of the most luminous sources in our list are RSGs. We find 556 objects (124 classified as O--rich, 17 as C--rich, and 415 as extreme AGB candidates) more luminous than M$_{\rm bol}=-7.1$, the classical AGB luminosity limit. It is not unlikely that some of these sources are AGB stars. Luminosities above the classical limit can be achieved by AGB stars at the peak of their pulsation cycle (as in the case of OH/IR stars) or by stars undergoing hot bottom burning \citep[HBB,][]{BoothroydSackmann1992}. Nevertheless, these numbers provide a very conservative upper limit for the RSG contamination in our sample.

The reddest sources in our list fall in the region of the MIR CMD also populated by YSOs. \citet{sage3} isolate regions in the [8.0] vs [8.0]--[24] CMD occupied more densely by YSO models (Figure 3 in their paper). This separation includes a stringent cut at [8.0]--[24]$\sim$2.2 (corresponding approximately to a 24 \mic\ flux $\ge$ the 8 \mic\ flux) to exclude AGB stars. We find that 5 O--rich and 24 extreme AGB star candidates in our list are part of the \citet{sage3} list of high-probability YSOs. A more conservative estimate is obtained by looking for sources fainter than [8.0]=7 with $F_{24 \mu m}/F_{8 \mu m}\ge 1$, which puts the YSO contamination in our lists at about a 100 sources (61 O--rich, 14 C--rich, and 34 extreme AGB candidates). 

\begin{figure}
\epsscale{0.8}\plotone{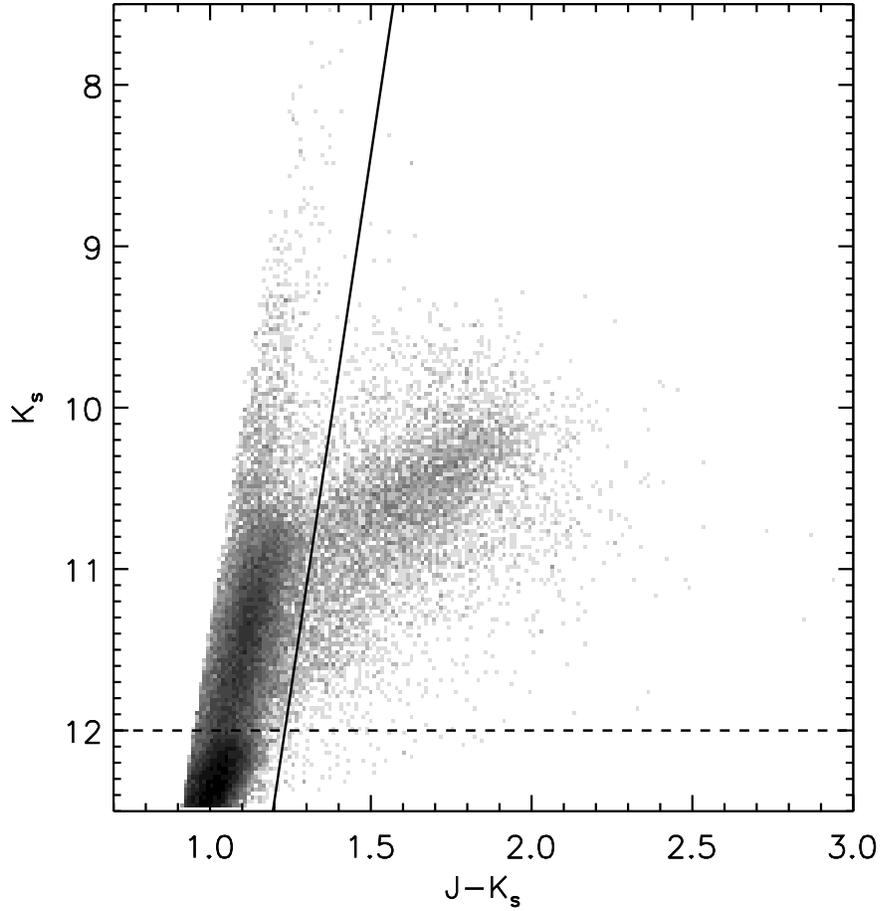}
\caption{A Hess diagram showing the locus of AGB stars in the \ks versus J--\ks CMD. Stars brighter than the tip of the RGB (K$_{\rm s} =$12, dashed line) are selected as AGB candidates, and these are classified as O--rich and C--rich based on the CMD cuts in \citet{Cioni}. The O--rich sources are bounded by the lines K$_1$ and K$_2$ (shown as solid line in figure) in their paper, while sources redward of K$_2$ are classified as C--rich. The extreme AGB candidates (not shown in this figure) are classified based on their 2MASS and IRAC colors.}
\label{jkcmd}
\end{figure}

\begin{figure}
\plotone{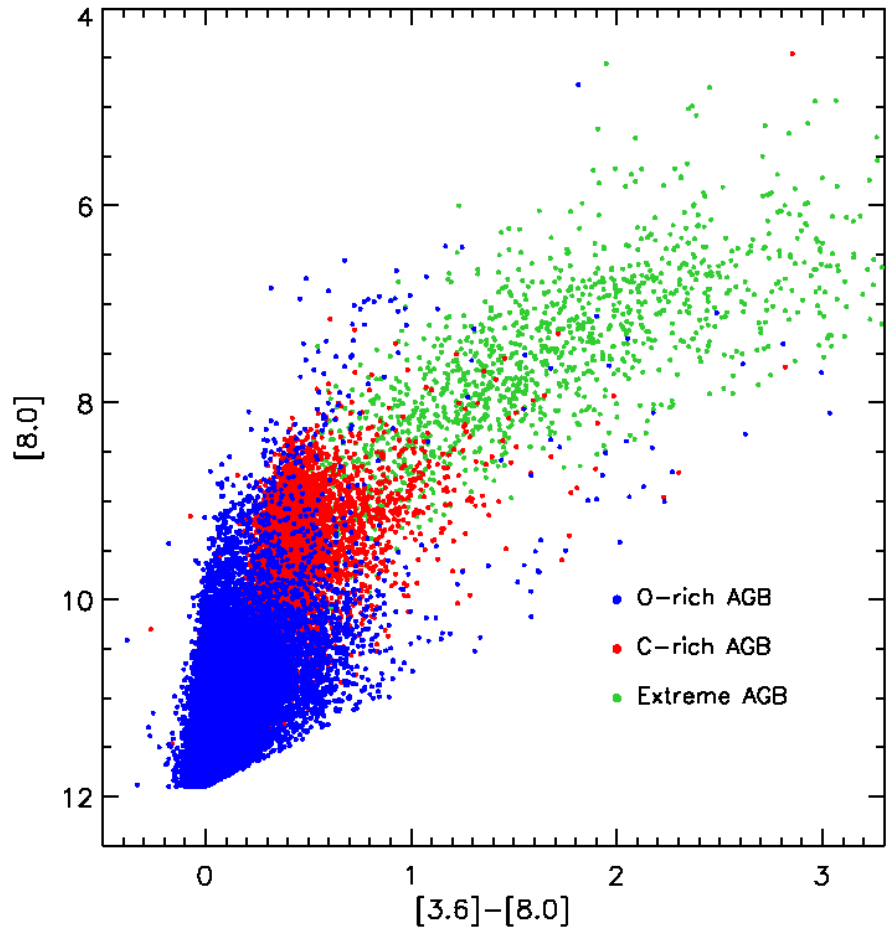}
\caption{The three types of AGB candidates (O--rich: blue, C--rich: red, Extreme: green) on an [8.0] vs [3.6]--[8.0] CMD. The tip of the RGB is at [3.6]$\approx$[8.0]=11.9.}
\label{iraccmd}
\end{figure}

\begin{figure}
\plotone{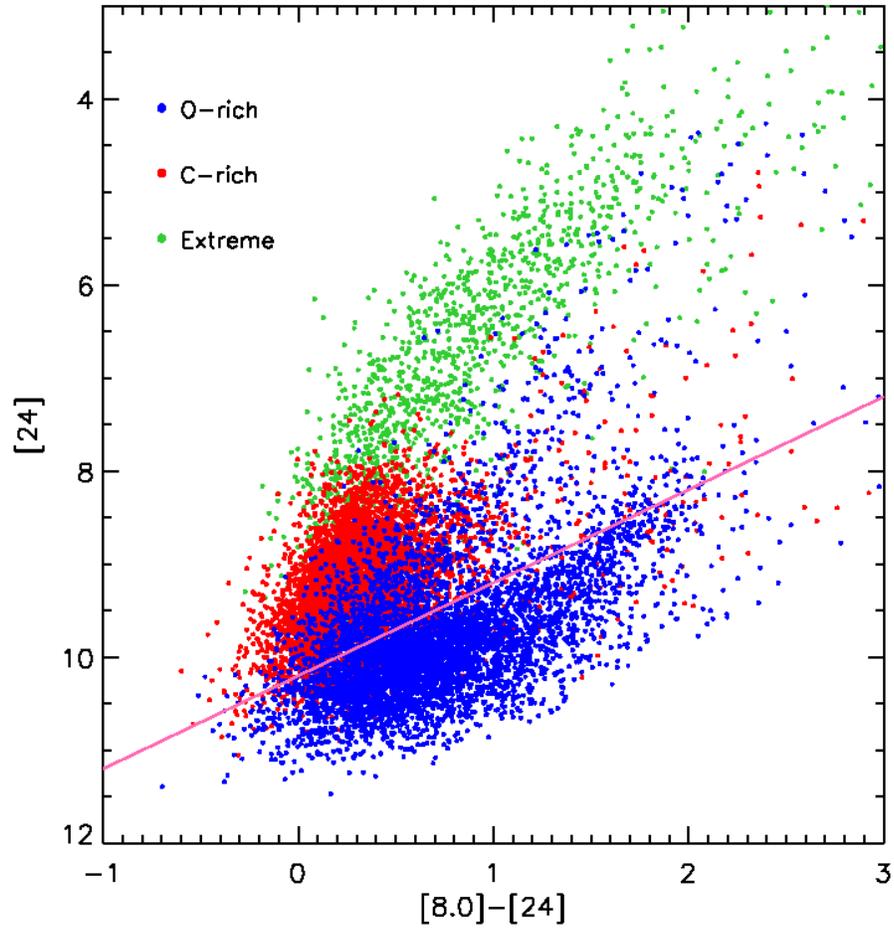}
\caption{The O--rich (blue), C--rich (red) and extreme AGB (green) stars on the [8.0]--[24] versus [24] CMD. The sources below the solid line are the fainter, redder population mentioned in \citet{sage2} (The finger ``F" in their Figure 6). About 80\% of the O--rich stars in our sample belong to this population.}
\label{twopops}
\end{figure}

Figure \ref{jkcmd} shows the \ks versus J-\ks Hess diagram for the region populated by AGB stars. The IRAC and MIPS24 CMDs are shown in Figures \ref{iraccmd} and \ref{twopops}. \citet{sage2} noted the presence of a fainter, redder population of O--rich sources (finger ``F'' in their Figure 6). The same population can be seen in our IRAC-MIPS24 CMD (Figure \ref{twopops}) at [8.0] magnitudes fainter than $\sim$10. We use a magnitude cut at 10.2, shown in the figure as a solid line. Throughout this paper, we will differentiate between the bright and faint O--rich populations based on this magnitude cut. Almost 80$\%$ of the O--rich stars in our sample are fainter than [8.0]=10.2. Photometric information for a few sources of each type is shown in Table \ref{SourceList}. The electronic version of this table contains information for our entire list of AGB star candidates.

\begin{figure}
\plotone{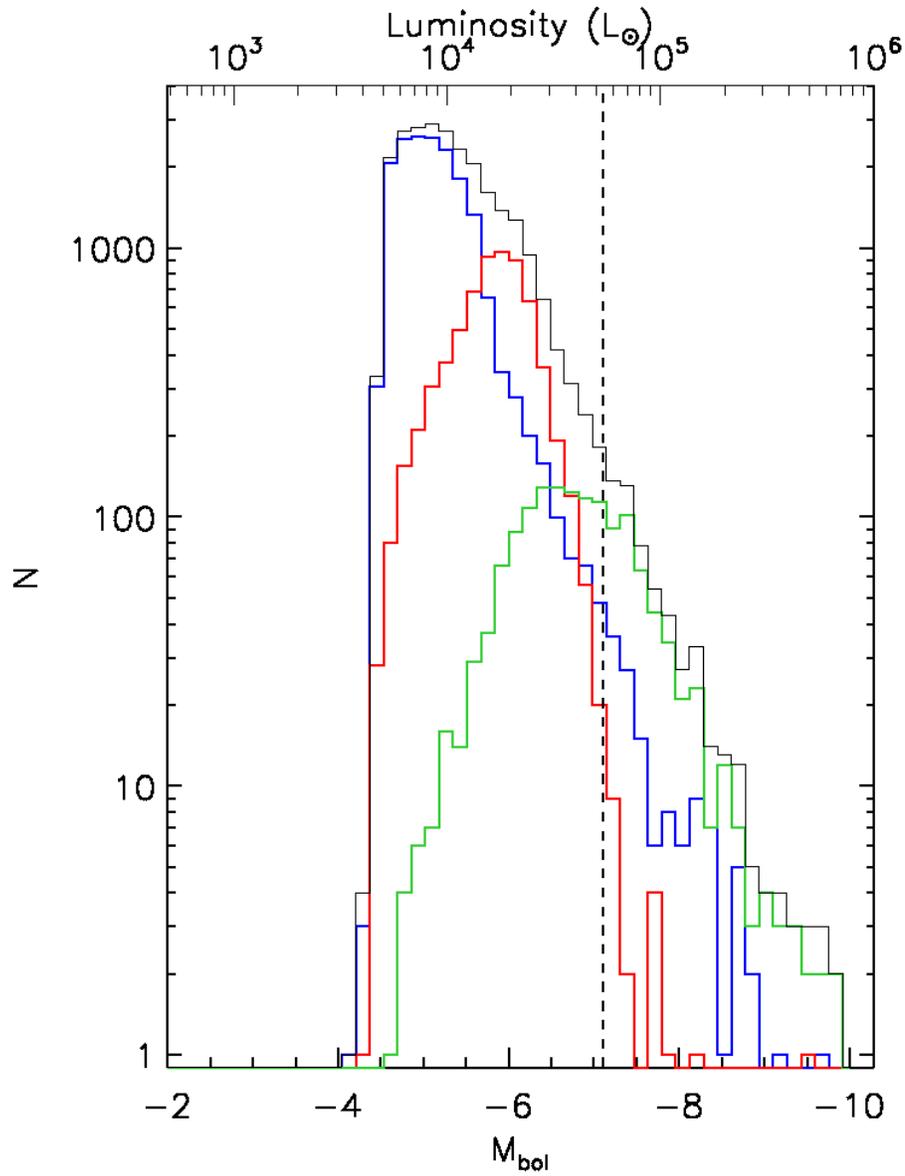}
\caption{The luminosity function for our sample of AGB candidates (blue: O--rich, red: C--rich, green: Extreme, black: total). The vertical dashed line is the classical AGB luminosity limit.}
\label{AGBLF}
\end{figure}

Figure \ref{AGBLF} shows the luminosity function for all three classes of AGB candidates. The lower luminosity limit ($M_{bol}\approx -4$ ) for the O--rich and C--rich AGB candidates is set by our color-magnitude cuts which exclude sources fainter than the tip of the RGB. The C--rich sources have a tighter luminosity distribution than their O--rich counterparts (there are only a handful of C--rich sources brighter than $M_{bol}=-7.5$, whereas the O--rich distribution falls off around $M_{bol}=-8.5$). The range in luminosities for the extreme AGB candidates is $M_{bol}=-4.5$ to $M_{bol}\approx -10$. The extreme AGB candidates thus have the highest luminosities in the sample. However, we have insufficient information at this point to say anything concrete about the breakdown of these sources into O--rich and C--rich chemistries. The spectroscopic follow-up to SAGE will provide some information about the dust chemistry of these extreme AGB candidates. These distributions peak at $M_{bol}\approx -5$ (O--rich), $M_{bol}\approx -6$ (C--rich) and $M_{bol}\approx -6.5$ (Extreme). The vertical dashed line shows the classical AGB luminosity limit. While this limit can be exceeded by deeply embedded AGB stars \citep{Woodetal1992}, O--rich sources with luminosities higher than $M_{bol}=-7.8$ are more likely to be RSGs, while the brightest extreme AGB stars may be massive YSOs. We will discuss the astrophysical implications of these luminosity functions in a future paper.

\section{Procedure}
\label{excess}
\noindent
We estimate the MIR excess emission of the low- and moderately-obscured O--rich and C--rich AGB star candidate populations by comparing their observed SEDs to an expected SED for the stellar photosphere. The IR excess in each wavelength band is calculated by comparing the total flux from the source observed in that band to the flux expected from the central star as prescribed by a ``best-fit" model photosphere. As the emission from CS dust dominates the MIR flux of extreme AGB stars, we set the MIR flux equal to the excess in each band for our extreme AGB star candidates. In this work, we are interested in describing overall trends in the AGB parameters such as IR excess as opposed to detailed models of each source. To this end we choose a single model photosphere for each type of AGB star -- one that best fits the SED shape of AGB stars with little or no dust. We use the plane-parallel C--rich MARCS models of \cite{Cmodels} and the spherical O--rich PHOENIX models of \cite{Omodels} to calculate the photospheric AGB star emission. The differences between plane-parallel and spherical models are insignificant as far as the resulting broad band AGB star photospheric information is concerned. The range of model parameters we search for a ``best-fit" are as follows: the thirty-two solar-metallicity C--rich AGB models have C/O ratios between 1.1 and 1.8, effective temperatures ranging from 2600 K to 3200 K, and surface gravity $\log{g}$ between about $-$0.76 and 0. The $\sim$200 O--rich AGB models of solar mass and solar metallicity have effective temperatures ranging from 2000 K to 4700 K, and $\log{g}$ between $-$0.5 and 2.5 in steps of 0.5. Synthetic photometry in the optical, near- and mid-infrared was obtained from each of these models by convolving their SEDs with the filter response curves of the Johnson-Kron-Cousins UBVI 
\footnote{The MCPS magnitudes were placed on the Johnson-Kron-Cousins UBVI system. The detector quantum efficiency curve was obtained from the Las Campanas Observatory website, 
http://www.lco.cl/lco/index.html. Filter profiles for the Johnson U, Harris B, V, and Cousins I filters were obtained from the references in Table 9 of \citet{Fukugita}.}
, 2MASS JH\ks
\footnote{The 2MASS filter relative
  spectral responses (RSRs) derived by \citet{2MASSRSRs} were obtained
  from the {\it 2MASS All-Sky Data Release Explanatory Supplement} on
  the worldwide web at http://www.ipac.caltech.edu/2mass/releases/allsky/doc/sec6$_{}$4a.html.}
, and Spitzer IRAC
\footnote{The IRAC RSRs are plotted in \citet{IRAC}, and were obtained from the {\it Spitzer Science Center} IRAC pages at http://ssc.spitzer.caltech.edu/irac/spectral$_{}$response.html}
 and MIPS 24 $\mu$m
\footnote{The MIPS \citep{MIPS} RSRs were obtained from the {\it Spitzer Science Center} MIPS pages at http://ssc.spitzer.caltech.edu/mips/spectral$_{}$response.html}
  filters. The oxygen-rich models had spectral information in the range 0.1--1000 \mic\ and the convolutions were done directly with the models. In order to compensate for the insufficient wavelength coverage ($\sim$0.5--25 \mic) offered by the carbon-rich models, the U and B band fluxes were dropped, and the flux in the MIPS24 band was extrapolated to 30 \mic\ assuming a Rayleigh-Jeans falloff.
  
\begin{figure}
\plotone{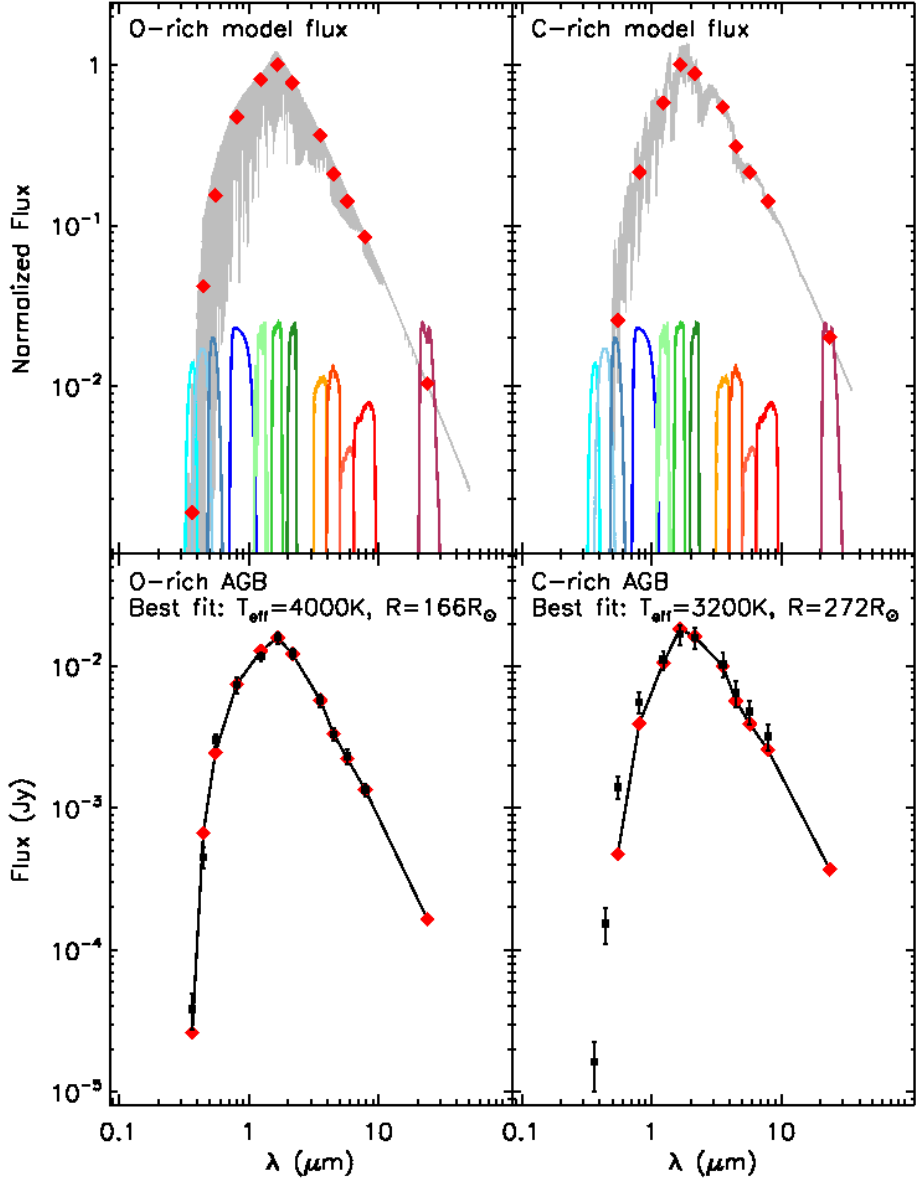}
\caption{The panels on the top show the normalized SEDs for the best-fit O--rich (left) and C--rich (right) models in light gray. The response curves for the optical UVBI, 2MASS JH\ks, IRAC and MIPS24 filters are also shown. Folding the transmission profiles of each filter into the model SED results in a flux in every filter (red diamonds). The bottom panels show the median flux in each band for the hundred bluest (in V--\ks color) O--rich (left) and C--rich (right) sources (black squares) plotted over the flux of the corresponding best-fit model (red diamonds) scaled to the median flux in the H band. The H band magnitudes of these bluest sources are in the range 11.99 to 11.36 (O--rich) and 11.95 to 10.32 (C--rich). The O--rich model is a 1 M$_\odot$ model with $T$=4000 K and $\log{g}=0$, while the C--rich model has $T$=3200 K, a C/O ratio of 1.3, and $\log{g}=-0.43$.}
\label{bestfitmodels}
\end{figure}

As we are most interested in obtaining the correct shape of the SED for a best-fit, the model SEDs are scaled to the H-band flux of the median SED of the hundred bluest sources in V$-$\ks color. We use the same model to calculate the excesses for both the bright and faint O--rich populations because the bluest O--rich candidates lack detections in the 24 \mic\ band and thus can not be separated into bright and faint populations. The model that comes closest to describing the oxygen--rich median SED has surface gravity $\log{g}$=0 and an effective temperature $T$=4000 K. The corresponding carbon--rich best-fit model parameters (for a C/O ratio of 1.3) are $\log{g}$=--0.43 and $T$=3200 K. Figure \ref{bestfitmodels} demonstrates the filter-folding for these best-fit models (top), and shows the comparison to the median SED of the bluest sources (bottom). There are many more O--rich candidates that are fainter and bluer than most of the C--rich candidates, which might explain the considerable difference in the best-fit model temperature and radius of the two types of sources.

To determine the excess flux $X_\nu$ in a band centered around frequency $\nu$ due to CS emission, the best-fit model flux must first be scaled to the flux of each source at some ``pivot" wavelength and then the difference between the observed flux $F_\nu$ and the corresponding scaled model flux in that band must be calculated. Fitting the SEDs of all our sources to one best-fit model photosphere is a simple first approach towards modeling the CSE around each AGB star candidate. The effects of interstellar and CS extinction on our sources is minimal in the 2MASS JH\ks bands, which also contain the wavelength range corresponding to maximum emission from AGB star photospheres.\footnote{The most obscured sources will suffer from CS extinction even in the NIR bands, but choice of pivot wavelength is not an issue for these sources, as they are probably members of our extreme AGB list.} Almost all of our O--rich sources peak in the H band\footnote{This is partly due to the fact that the opacity of the H$^-$ ion reaches a minimum in the H-band.}, while over two-thirds of the C--rich sources peak in the \ks band. All of the photospheric models available to us exhibit H-band SED maxima. For these reasons, we place the pivot wavelength in the H band (centered at 1.65 \mic). The excess then depends on the source flux $F_\nu$ and model flux $F^{mod}_\nu$ according to
\begin{eqnarray}
X_\nu=F_\nu-\left(\frac{F^{mod}_\nu}{F^{mod}_{\rm H}}\right)F_{\rm H}
\end{eqnarray}
where the subscript H denotes H-band fluxes. This equation may overestimate the MIR excesses from the redder C--rich sources with $F_{K}> F_{H}$. On the other hand, using the same best-fit model with a \kss-band pivot would produce underestimates for the excesses. For the reddest C--rich source in our sample, the difference in 8 \mic\ excess resulting from choice of pivot is $\sim 30\%$. This is to be regarded as an upper bound to the error introduced in the excess determination due to the choice of the H-band as pivot. While this will not alter the general trends we discuss in this paper, it will affect our numerical results for higher excesses (mass-loss rates).

\begin{figure}
\epsscale{0.8}\plotone{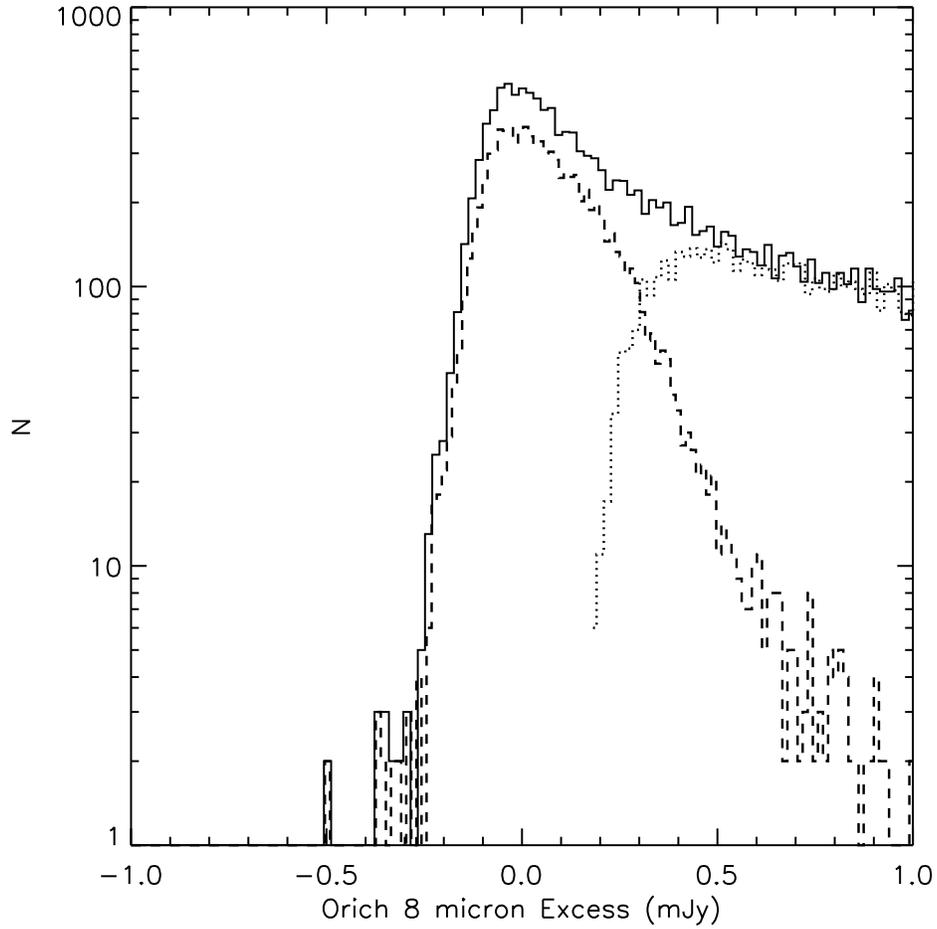}
\caption{Histogram of the 8 \mic\ excesses for the faintest O--rich AGB candidates (and thus with the lowest S/N ratio excess emission). The solid line is for all sources. The dotted line shows sources with S/N$>$3. The dashed line shows sources with excesses considered unreliable and hence were excluded from our study.} 
\label{xshisto}
\end{figure}

The photometric errors associated with the source fluxes were used to estimate the error $\delta X$ in the calculated excess. An excess measurement in a wavelength band centered around frequency $\nu$ was deemed ``reliable" only if its signal-to-noise ratio was greater than 3. In other words,
\begin{eqnarray}
\label{qualcut}
{\delta X_\nu \over X_\nu} \leq {1 \over 3}
\end{eqnarray}
Figure \ref{xshisto} illustrates the effect of selecting sources using this ``3-$\sigma$" criterion. The distribution of sources that are rejected based on this criterion is fairly symmetric about zero, with a slight asymmetry on the positive excess side, indicating that our cut is conservative. A similar distribution of rejected sources is seen for the C--rich candidates, but they are substantially fewer in number. In both cases, sources with excesses below $\sim$ 0.1 mJy are rejected.

Very little can be inferred about the chemical composition of the dust shells around the extreme sources without spectroscopic confirmation, although most of these objects are probably C--rich. We find that 60 of our extreme AGB candidates are identified as C--rich in the \citet{Kastneretal2008} study, while only 9 are classified as O--rich. Our procedure for calculating excesses relies on a classification into O--rich or C--rich sources, which is not possible for most of the stars in this list. However, the excess emission from their extremely dusty CSEs dominates over the photospheric emission in the MIR, so that we can set the excess equal to the MIR flux to a good level of approximation. We find $\sim$ 8200 O--rich, 5800 C--rich and 1400 extreme sources with reliable 8 \mic\ excesses, and about 4700 O--rich, 4900 C--rich and 1300 extreme sources with reliable 24 \mic\ excesses. Table \ref{ExcessList} shows the excesses calculated for the fifteen sample sources shown in Table \ref{SourceList}. The electronic version of Table \ref{ExcessList} lists all the sources with valid 8 \mic\ excesses.

We estimate the temperature of the CS dust and the optical depth from the 8 \mic\ and 24 \mic\ excesses in a manner similar to \citet{Thompsonetal2006} and \citet{Dayaletal1998}. The continuum dust emission is modeled as a blackbody at temperature $T_d$, with optical depth $\tau_\nu$,
\ben
\label{coltau}
X_\nu \propto \Omega_\nu (1-e^{-\tau_\nu}) B_\nu(\lambda, T_d) 
\een
where 
\ben
\Omega_\nu=\pi\left(\frac{R_{in}}{D}\right)^2
\een
is the geometrical dilution factor with $R_{in}$ the distance from the central star corresponding to maximum emission due to dust at both 8 and 24 \mic. For small optical depths, the excess is proportional to the opacity $\kappa_\nu$. If the emissivity of the dust can be modeled by a $\lambda^{-1}$ power law in the relevant range of wavelengths, we have
\ben
\label{coltemp}
X_\nu \propto \Omega_\nu\lambda^{-1}B_\nu(\lambda, T_d)
\een

Assuming a power-law emissivity in the MIR ignores any effects due to strong absorption or emission features from silicate dust for example, which can be significant for the more obscured O--rich sources. The power-law index will also depend on the dust species in general \citep[see, {\it e.g.},][]{Hofner2007}. The  $\lambda^{-1}$ power-law dependence adequately describes the emissivity at MIR wavelengths for the purpose of this paper (the {\it Spitzer} 8 and 24 \mic\ bands will only detect the wings of silicate emission unless the sources are OH/IR stars.) The color temperature $T_d$ is calculated by constructing a look-up table with the excess ratio calculated as per Equation \ref{coltemp} for a wide range of temperatures, and selecting the value of temperature from this table that reproduces the observed ratio of excesses. Once the color temperature is calculated, Equation \ref{coltau} can be solved for the optical depth. The color temperature and optical depth values calculated using these simplified equations are primarily useful in investigating the trends of color temperature and optical depth with excess. The equations break down for sources with high optical depth ({\it i.e.}, for extreme AGB stars). The temperature derived from a ratio of two broad band fluxes is a simplification that will in turn affect the optical depth calculation.  We will model the CS shells of our AGB candidates considering details such as the wavelength dependence of the emissivity and the choice of dust species in a future paper to obtain more precise estimates for the temperatures and optical depths.

Since we are only interested in the overall trend of the optical depth with excess flux, we fix $R_{in}$ at ten times the stellar radius $R_*$, which is calculated by scaling the radii of the model photospheres to the luminosity of each star. According to models of dust condensation, most of the dust has condensed by $\sim$ 10$R_*$ \cite[see, {\it e.g.},][]{Hofner2007}. In practice, the dust condensation radii and $R_{in}/R_*$ values will vary, but this complication is ignored in this work as we are only interested in a comparative study of the MIR color temperatures and optical depths. While the temperature of CS dust varies with radius from $\sim$1000 to $\sim$100 K, the (single) color temperature calculated here will be dominated by the distance corresponding to the hottest and most optically thick dust emission in the CS shell, which is typically close to the inner radius of the dust shell. Our values for $\tau_\nu$ describe the optical depth of the high density inner region of the CS shell. This estimate is insensitive to large but cool optical depths. This optical depth will be the dominant component of the line of sight optical depth. Table \ref{TempTauList} shows the color temperatures and optical depths for the sources in Tables \ref{SourceList} and \ref{ExcessList}. Data for the entire set of sources with valid 8 \mic\ excesses is available in the electronic version of Table \ref{TempTauList}.

\section{Results}
\label{results}
\noindent In this section, we investigate the relation of the MIR excess to other observed parameters for the sample including luminosity, color temperature of the CS dust, and opacity of the CS dust shell. We focus on the statistical nature of the sample and global trends of the results. This approach provides the range of physical parameters for the sample and is a first and necessary step to obtain appropriate ranges for model parameters for any detailed modeling that may follow for the sample.

\subsection{Excess-Luminosity Relations}
\begin{figure}
\epsscale{0.8}\plotone{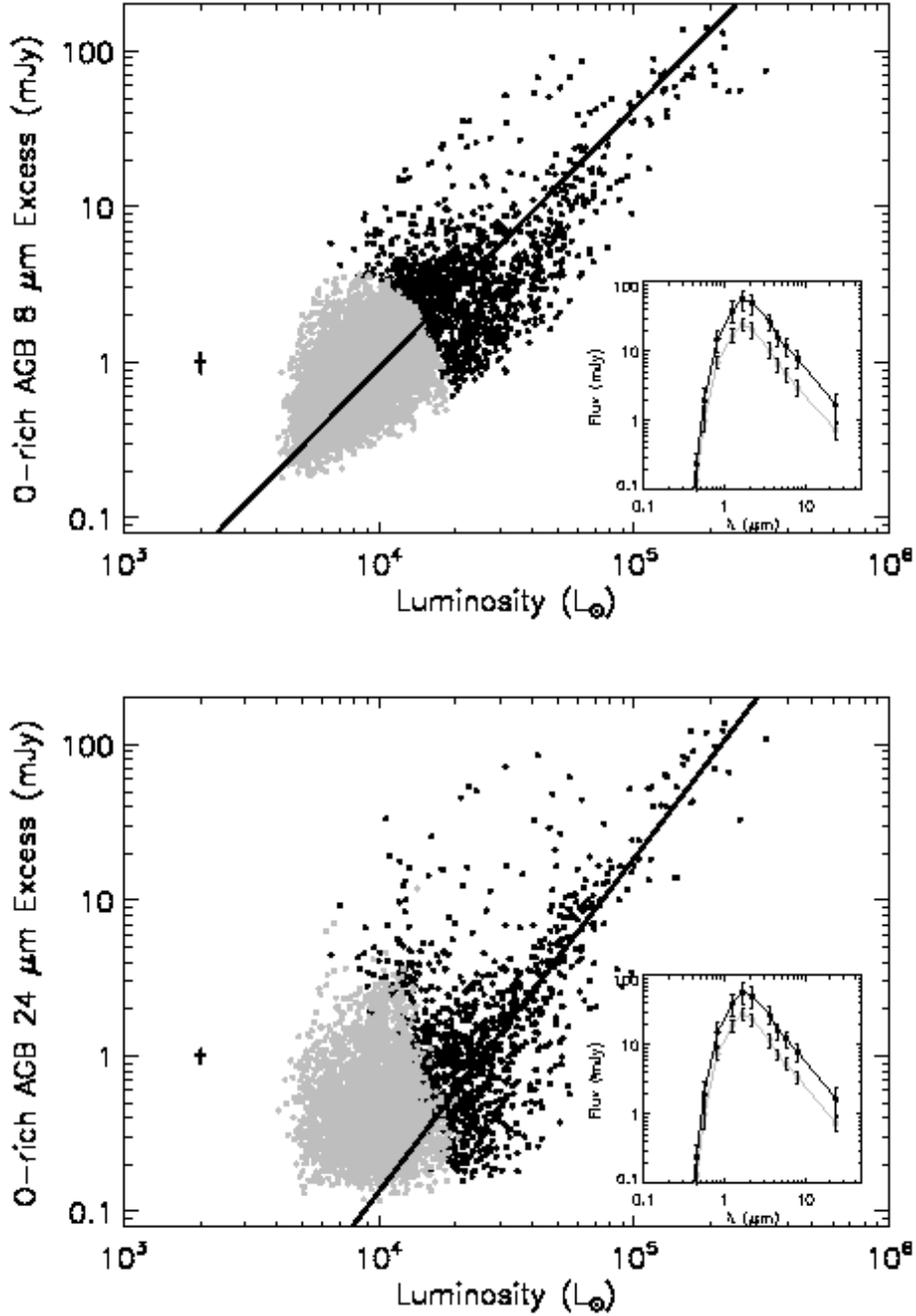}
\caption{The 8 \mic\ (top) and 24 \mic\ (bottom) excesses plotted against luminosity for the O--rich sources. Sources from the bright and faint populations (light gray and dark gray circles respectively) are shown.  The solid lines show, respectively, power-law fits to all the sources with reliable 8 \mic\ excesses (top) and sources with reliable 24 \mic\ excesses brighter than $3\times 10^4 $ L$_\odot$ (bottom). Representative error bars are also shown. Inset: the median SEDs of the bright and faint populations.}
\label{Oxs}
\end{figure}

\begin{figure}
\epsscale{0.8}\plotone{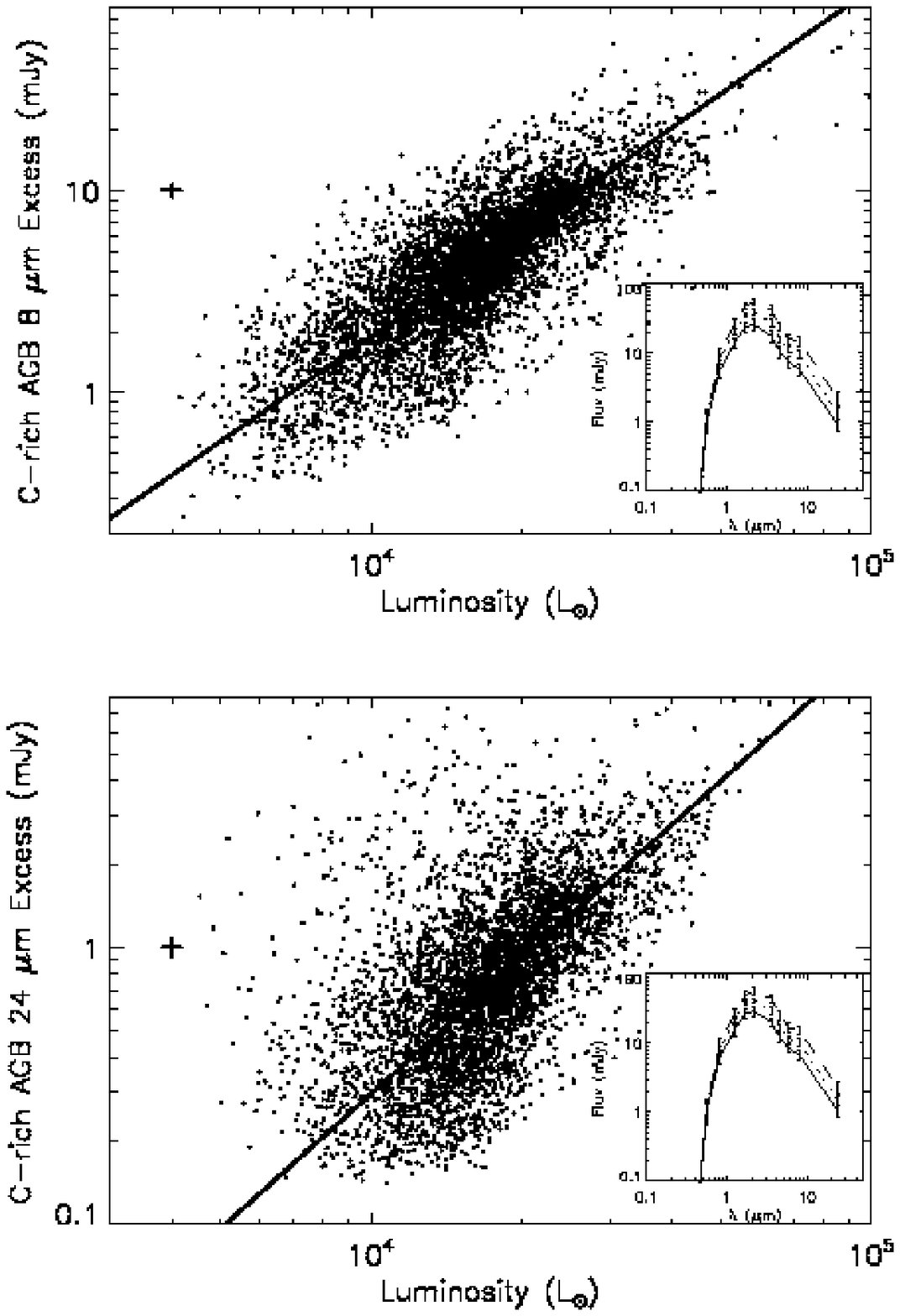}
\caption{The 8 \mic\ (top) and 24 \mic\ (bottom) excesses plotted against luminosity for the C-rich sources. The solid lines are power-law fits to the excess--luminosity relations. (The fit is very poor in the 24 \mic\ case) Representative error bars are also shown. Inset: the median SEDs for three equally-populated luminosity bins (solid line: least luminous one-third of the sample, thin dashed line: intermediate luminosity, thick dashed line: most luminous).}
\label{Cxs}
\end{figure}

\begin{figure}
\epsscale{0.8}\plotone{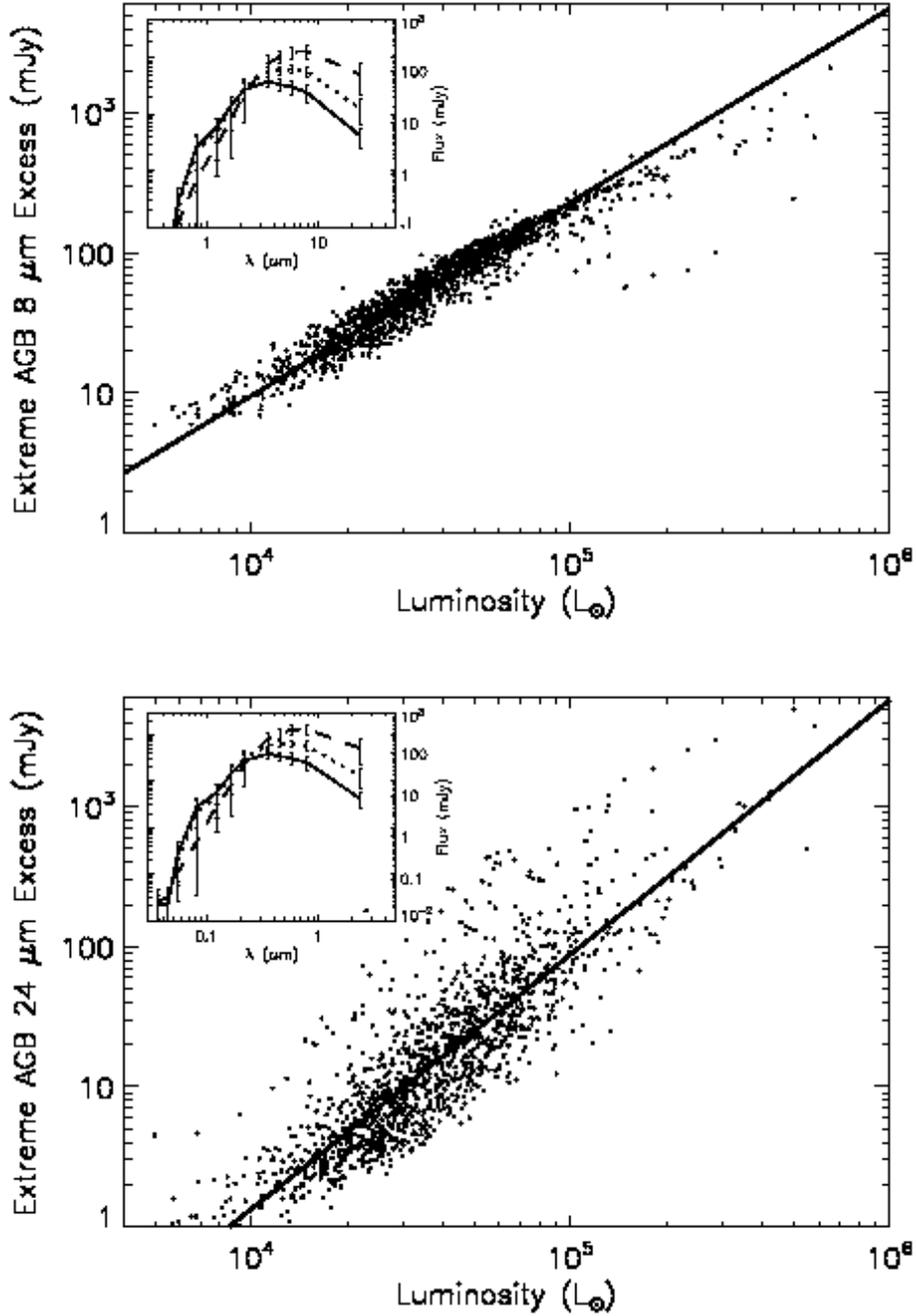}
\caption{The 8 \mic\ (top) and 24 \mic\ (bottom) excesses plotted against luminosity for the extreme sources. The solid lines are power-law fits to the excess--luminosity relations. Representative error bars are also shown. Inset: the median SEDs for three equally-populated luminosity bins (solid line: least luminous third of sample, thin dashed line: intermediate luminosity, thick dashed line: most luminous).}
\label{Xxs}
\end{figure}

\noindent The calculated IR excesses show a general increasing trend with luminosity at all IRAC wavelengths for O-- and C--rich AGB sources. This trend is most obvious at 8 \mic\ (Figures \ref{Oxs} and \ref{Cxs}). While the correlation is similar at the shorter IRAC wavelengths, the number of sources with significant reliable excesses is considerably smaller. This is expected for the warmer, moderately obscured non-extreme sources. The 8 and 24 \mic\ excesses of the extreme AGB candidates also correlate well with their luminosities, as is evident from Figure \ref{Xxs}. There is a huge spread in 24 \mic\ excesses for the O--rich and C--rich candidates overall, but an increasing trend is apparent for the O--rich AGB candidates at luminosities above $\sim 3\times 10^4$ L$_\odot$. The large spread at low luminosities may be caused by significant variations in the MLRs during the early stages of AGB stars. The plots of O--rich sources show the bright and faint population sources in light and dark grey respectively. We also display the median SED for each population as an inset. For the C--rich and extreme sources, we show the median SED for three equally-populated bins in luminosity. Relations of the form
\ben
\label{powerlaw}
\log{X_\nu({\rm mJy})}=A+B\log{L ({\rm L}_\odot)}
\een
are fit to the 8 and 24 \mic\ excesses, resulting in the following best-fit parameters:
\ben
\nonumber
A=-6.7\pm 0.5,\ B=1.7\pm 0.1~\rm{(O-rich, 8\ \mu m)}\\
\nonumber
A=-9.4\pm 3.2,\ B=2.1\pm 0.7~\rm{(O-rich, 24\ \mu m, L> 3\times 10^4~L_\odot)}\\
\nonumber
A=-6.6\pm 6.2,\ B=1.7\pm 1.5~\rm{(C-rich, 8\ \mu m)}\\
\nonumber
A=-7.0\pm 15, \ B=1.6\pm 3.5~\rm{(C-rich, 24\ \mu m)}\\
\nonumber
A=-5.0\pm 0.3,\ B=1.5\pm 0.1~\rm{(Extreme, 8\ \mu m)}\\
\nonumber
A=-8.3\pm 0.7,\ B=2.1\pm 0.2~\rm{( Extreme, 24\ \mu m)}
\een
The poor fit to the 24 \mic\ excess for the C--rich candidates reflects the huge spread of 24 \mic\ excesses for these sources.

\subsection{Color Temperature and Opacity}
\begin{figure}
\epsscale{0.8}\plotone{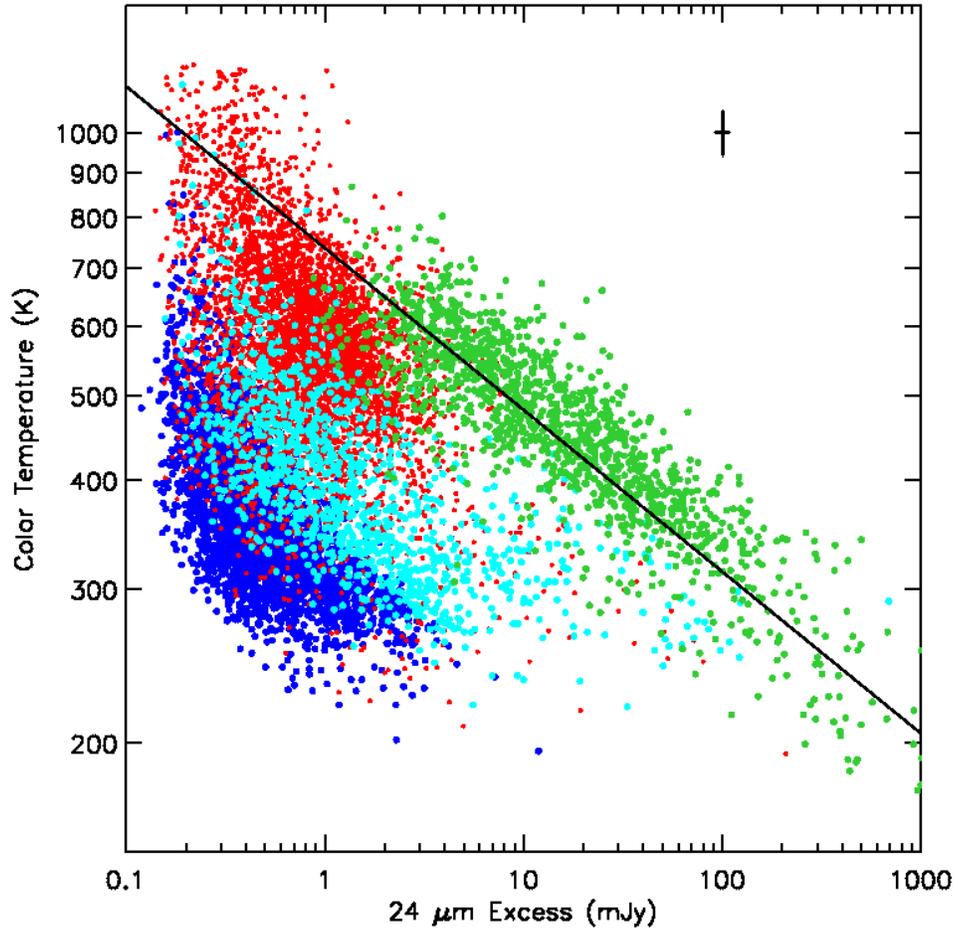}
\caption{The color temperature derived from the ratio of the 8 \mic\ and 24 \mic\ excesses plotted against the 24 \mic\ excess for all three types of AGB candidates (blue: faint O--rich, cyan: bright O--rich, red: C--rich, green: Extreme). The cross is a representative error bar.}
\label{temp}
\end{figure}

\begin{figure}
\epsscale{0.8}\plotone{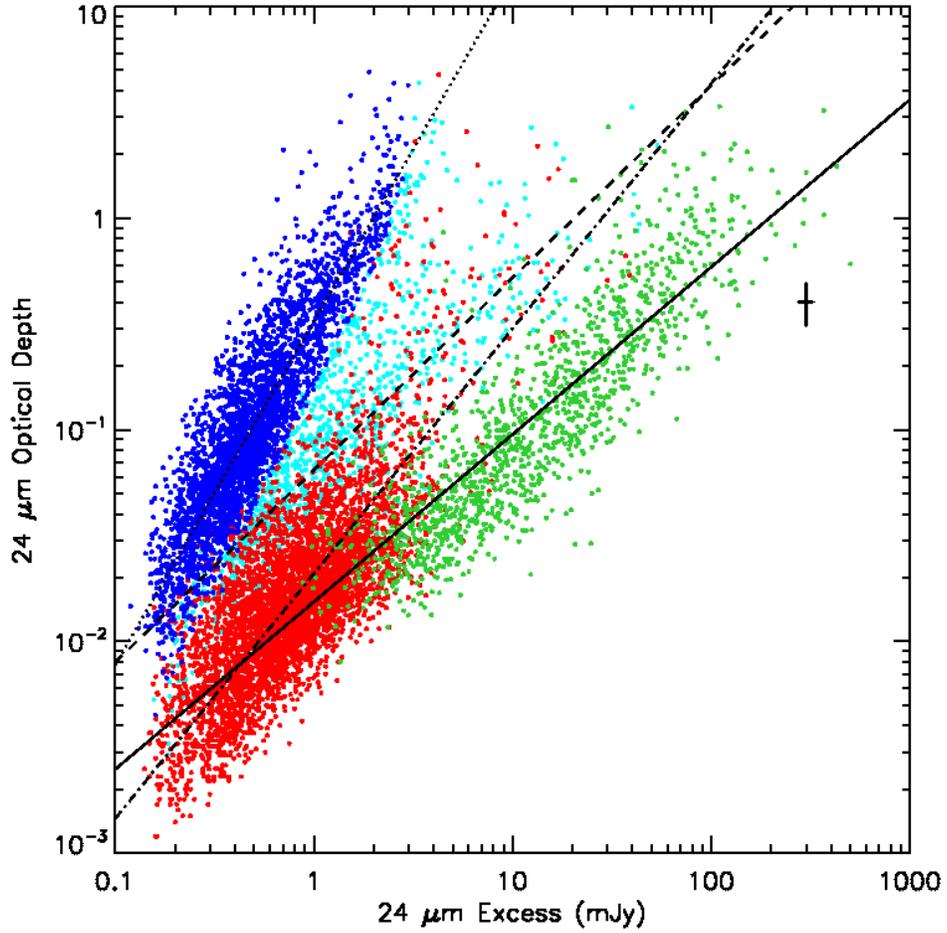}
\caption{The 24 \mic\ optical depth derived from the 8 \mic\ and 24 \mic\ excesses for all three types of AGB candidates. The color coding is the same as for Figure \ref{temp}.  Power-law fits to the optical depth--excess relation are shown (dotted: faint O--rich, dashed: bright O--rich, dot-dashed: C--rich, solid: Extreme). The cross is a representative error bar.}
\label{tau}
\end{figure}

\begin{figure}
\epsscale{0.8}\plotone{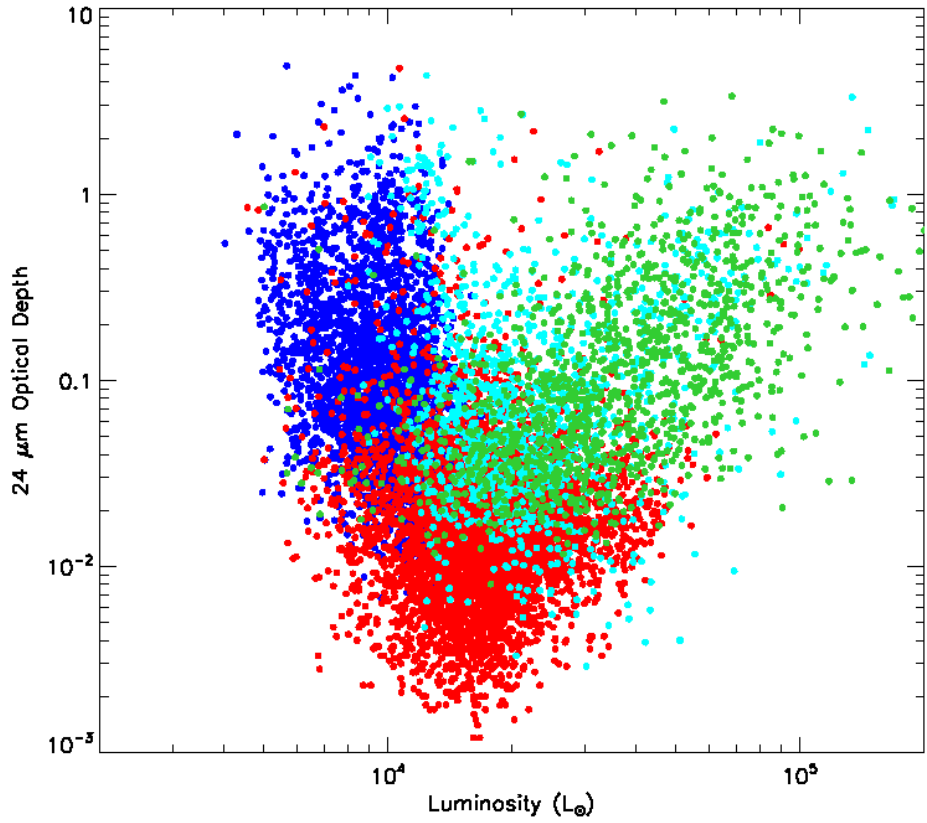}
\caption{The 24 \mic\ optical depth as a function of luminosity for all three types of AGB candidates. The color coding is the same as for Figure \ref{temp}.}
\label{tauvsl}
\end{figure}

\begin{figure}
\epsscale{0.8}\plotone{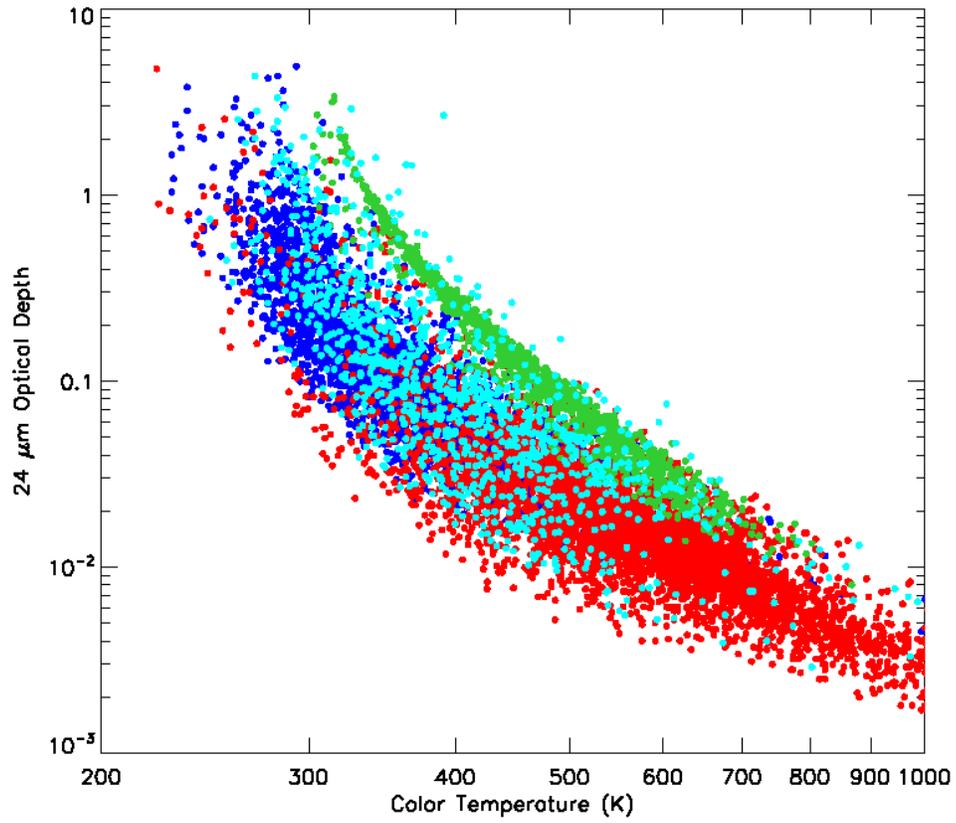}
\caption{The 24 \mic\ optical depth plotted against color temperature for all three AGB types. The color coding is the same as for Figure \ref{temp}.}
\label{tauvstemp}
\end{figure}

\noindent Figures \ref{temp} and \ref{tau} show the variation of the color temperature and 24 \mic\ optical depth with 24 \mic\ excess for all three types of AGB candidates. As expected from their redder [8]$-$[24] color, the members of the faint O--rich population tend to have cooler temperatures and higher optical depths than the bright population, and they also show a more pronounced variation with excess. There is a huge spread in the C--rich AGB color temperatures. The maximum color temperature is higher than in the O--rich case, suggesting that the carbonaceous dust grains become hotter than the silicates. AGB winds composed of oxygen-rich compounds are less efficient at absorbing visible photons than those that are carbon-rich \citep{WK1998}. Carbonaceous grains are efficient absorbers of optical photons and are highly emissive at IR wavelengths in comparison to silicates, thus they are more likely to reach higher temperatures. We fit the color temperatures for the extreme AGB candidates with a power-law relation of the form
\ben
\nonumber
\log{T_d({\rm K})}=A+B\log{X_\nu({\rm mJy})} 
\een
and obtain the following best-fit values:
\ben
\nonumber
A=2.9\pm 0.4,\ B=-0.19\pm 0.02~\rm{(Extreme)}
\een
while a similar relation,
\ben
\log{\tau_{24 \mu {\rm m}}}=A+B\log{X_\nu({\rm mJy})} 
\een
when fit to the 24 \mic\ optical depths, gives
\ben
\nonumber
A=-0.49\pm 0.12,\ B=1.6\pm 0.1~\rm{(O-rich, faint)}\\
\nonumber
A=-1.2\pm 0.3,\ B=0.9\pm 0.2~\rm{(O-rich, bright)}\\
\nonumber
A=-1.7\pm 0.3,\ B=1.2\pm 0.2~\rm{(C-rich)}\\
\nonumber
A=-1.8\pm 0.1,\ B=0.79\pm 0.09~\rm{(Extreme)}
\een
Figures \ref{tauvsl} and \ref{tauvstemp} show the variation of the 24 \mic\ optical depth with luminosity and color temperature respectively. There is no trend in the optical depth of the O--rich candidates with luminosity. For the C--rich sources, there is no correlation, but there appears to be an absence of sources with low optical depth at higher luminosities. For the extreme AGB stars, there appears to be a positive correlation of higher optical depth with higher luminosities. The optical depth decreases monotonically with the color temperature for all three types of AGB candidates. CS dust shells have temperature gradients from $\sim$ 1000 K in the interior to $\ltapp$ 100 K in the outer regions. With increasing optical depth, therefore, we are probing the progressively cooler, outer regions of the AGB star candidates.

\section{Discussion}
\label{discuss}
\noindent The 8 and 24 $\mu$m band excess fluxes show an increase with increasing luminosity of the central star. For each type of AGB candidate, the slope of the excess-luminosity relation is similar across the four IRAC and MIPS24 bands, though in the IRAC 8 \mic\ and MIPS24 bands, excesses are ``reliable" only beyond $\sim$0.1 mJy (see Figure \ref{xshisto}). The increase of excess with luminosity is consistent with the MLR--luminosity relation found by \citet{vL1999}. The mass loss-rate is roughly proportionate to $\tau$$L$ \citep[see, {\it e.g.},][]{IE1995}, so we expect that the MLR increases with increasing luminosity as long as the optical depth does not decrease faster than $\sim$L$^{-1}$. This suggests that the excess is a good reflection of the MLR. More quantitative comparisons of MLR for this whole sample with luminosity is beyond the scope of this study, but this will be addressed in a future paper.

The uncertainties in photometry alone cannot account for the considerable spread in the calculated excesses. Stars on the AGB suffer from variability and episodic mass loss, and our observations of these stars may be capturing their fluxes at different epochs in their variability cycle, or during episodes of increased or diminished mass loss. At a given luminosity, these effects introduce a variation in the MLR and hence also affect the excess. \citet{Vijhetal} show a median variability index of about 3 for the O--rich and C--rich AGB sources and about 5 for the extreme sources. This would correspond to a change in the excess fluxes of about the same factor. The color temperature depends on the ratio of the excesses, and the effect of variability is probably weaker. The effects of grain chemistry, metallicity of the environment, and the C/O ratio of these sources have not been accounted for in this study, as this requires spectroscopic determination. At a given luminosity, there is also a dependence on progenitor mass -- a degeneracy between a less evolved, more massive star and a more evolved, less massive star arises. Such degeneracy effects increase the spread at the lower luminosity end (only the most massive stars evolve to high luminosities).

Observations for Galactic AGB stars \citep{Guandalinietal06,Bussoetal07} show that C--rich AGB stars are in general more obscured than their O--rich counterparts, but this is partly due to the higher opacity of carbonaceous dust. Note that extreme O--rich AGB stars ({\it i.e.}, OH/IR stars) are even more obscured than these C--rich sources. In the case of the extreme sources, there is a high optical depth regardless of chemistry of the CSE. As an AGB star evolves, more mass in the form of dust is deposited into the surrounding shell, increasing its opacity and causing further reddening of stellar light and stronger emission at longer wavelengths, corresponding to cooler color temperatures. The color temperature and optical depth variations for extreme AGB stars in Figures \ref{temp} and \ref{tau} support these claims. The existence of a redder, fainter population of O--rich sources first seen in the IRAC CMD is also apparent in plots of the excess, color temperature and optical depth. The faint population is in general cooler and more obscured than the bright population. The SAGE survey is able to differentiate between these two types of sources for the first time. The bright population corresponds to the young, most massive AGB stars that prevent the dredge-up of carbon by undergoing hot-bottom burning (part of sequence G in \citet{2MASSLMC}). This population also emerges in the model isochrones of \citet{Marigoetal} -- they find that the thermally-pulsing (TP) phase in their isochrone of age $\log({t/{\rm yr}})=8.2$ is well-developed and populated by O--rich stars undergoing HBB. The faint population consists of AGB stars in which HBB does not occur, resulting in a smooth transition from O--rich to C--rich chemistry. The $\log({t/{\rm yr}})=9.1$ isochrone of \citet{Marigoetal} (sequence J in \citet{2MASSLMC}) represents the typical evolutionary phase corresponding to this population. 

\begin{figure}
\epsscale{0.8}\plotone{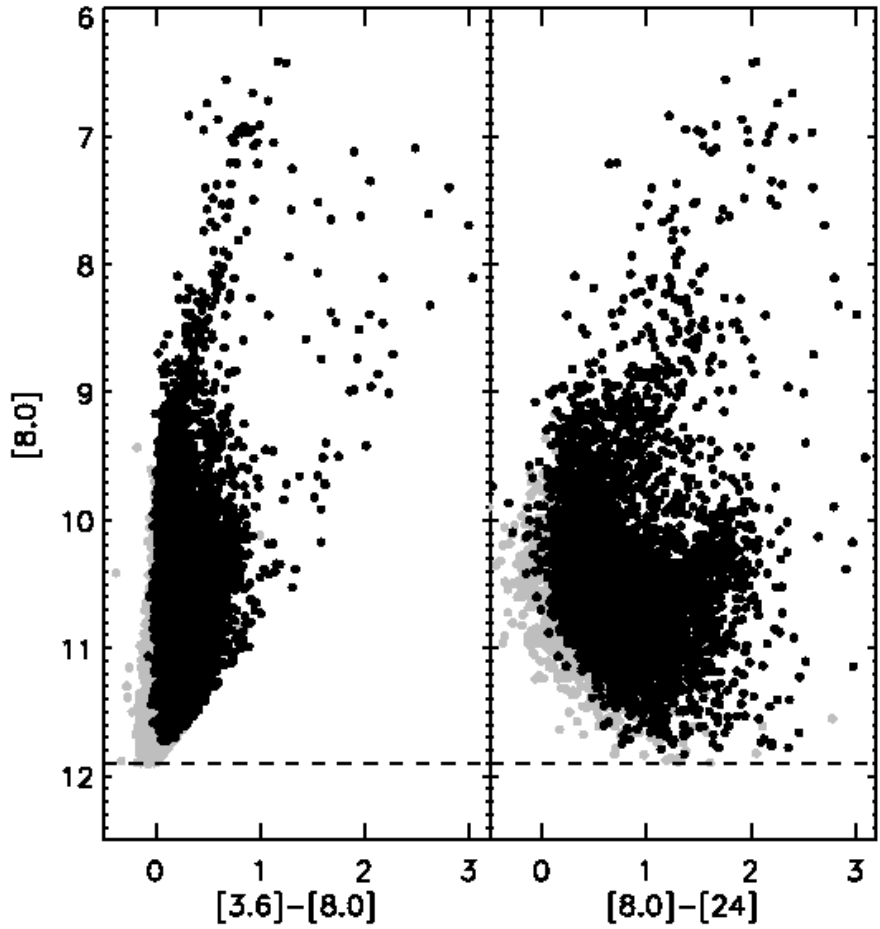}
\caption{[8.0] vs [3.6]--[8.0] and [8.0] vs [8.0]--[24] CMDs for the O--rich AGB stars in our list with 8 and 24 $\mu$m detections (gray circles). The sources with ``reliable'' excesses are superimposed (black circles). The tip of the RGB is at [8.0]=11.9 (dashed line)}
\label{discspace_Orich}
\end{figure}

\begin{figure}
\epsscale{0.8}\plotone{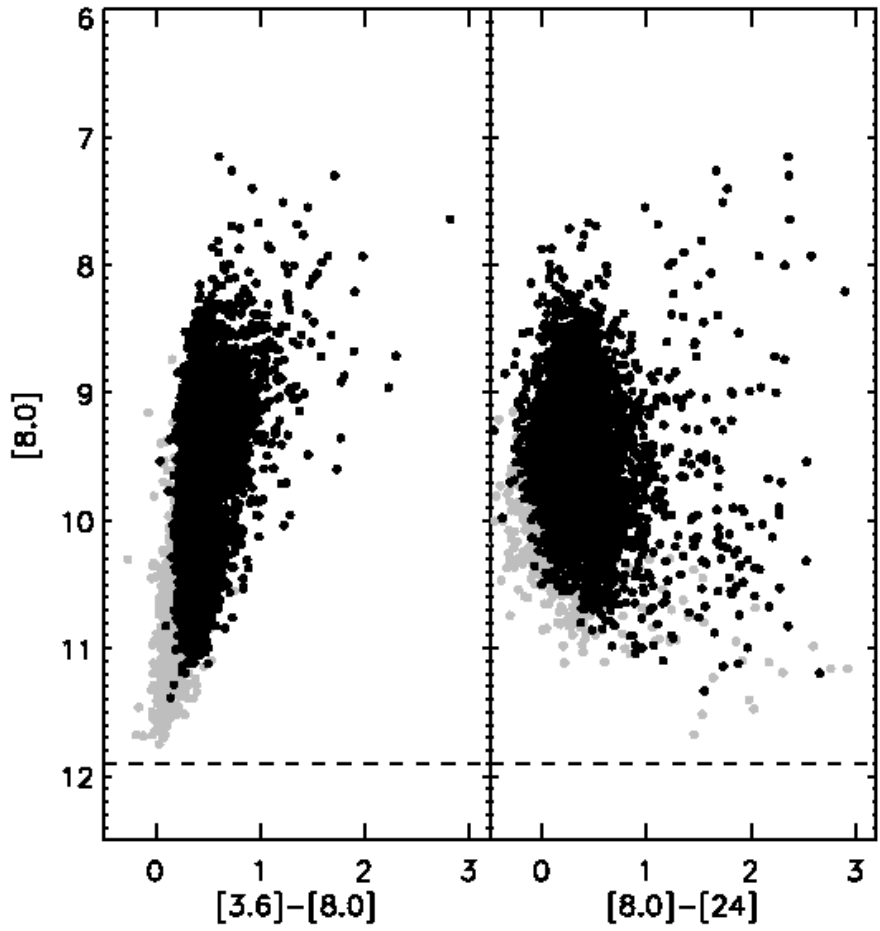}
\caption{Same as Figure \ref{discspace_Orich} for our C--rich AGB candidates}
\label{discspace_Crich}
\end{figure}

\begin{figure}
\epsscale{0.8}\plotone{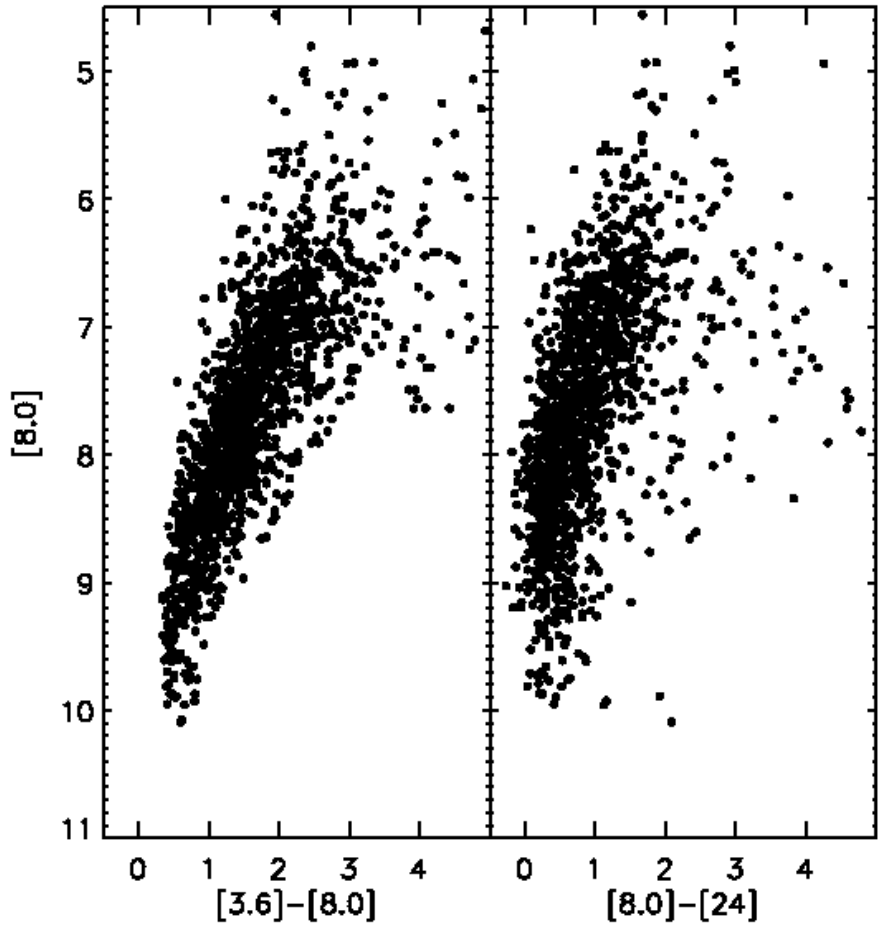}
\caption{Same as Figure \ref{discspace_Orich} for our extreme AGB candidates}
\label{discspace_xAGB}
\end{figure}

Our $\sigma$-clipping method of recovering sources with reliable excesses can be used to compare our results with the expected lower-limit to measurement of MLRs by the SAGE survey \citep{sage1}. The mid-infrared CMDs in Figures \ref{discspace_Orich}, \ref{discspace_Crich}, and \ref{discspace_xAGB} show the sources with reliable 8 \mic\ excesses as black circles. We detect O--rich and C--rich sources with reliable excesses up to within $\sim$ 0.1 mag of the tip of the RGB ([8.0]$\approx$11.9). which is below the detection limit mentioned in \citet{sage1} ([8.0]=11.0) for measuring significant mass loss. The extreme AGB stars in our sample are all well above this detection limit, showing that the SAGE survey detects all the extreme mass-losing AGB sources in the LMC. 

\begin{figure}
\epsscale{0.8}\plotone{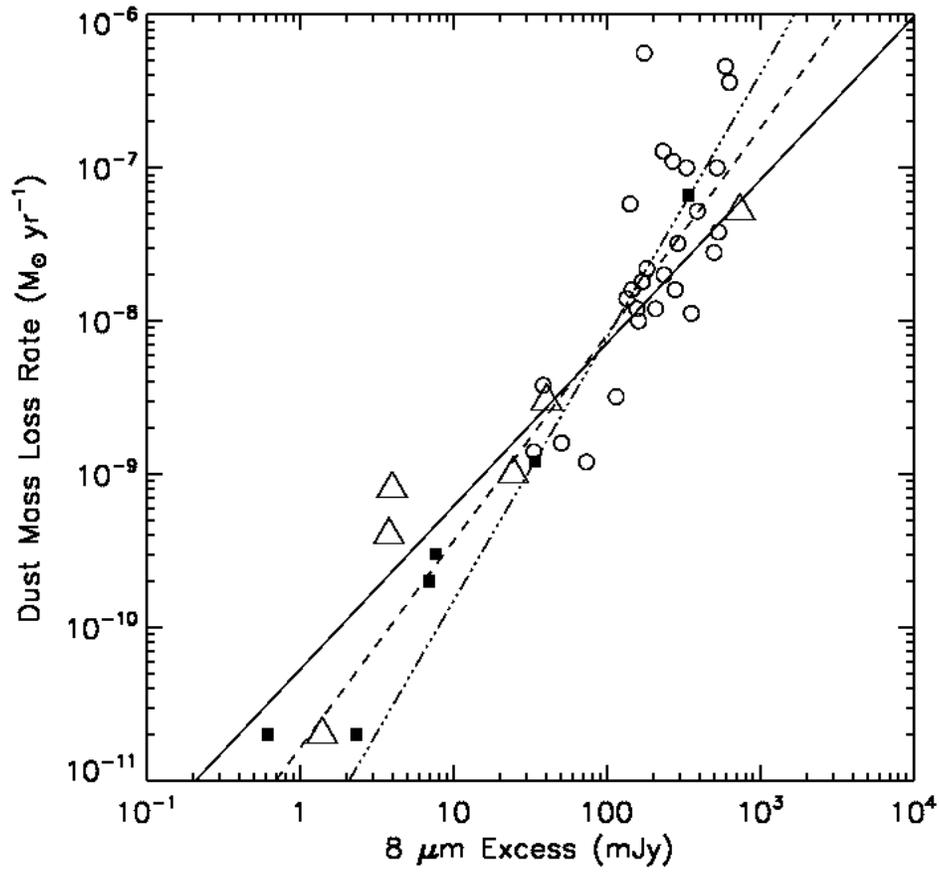}
\caption{The mass-loss rates from \citet{vL1999} for O--rich (triangles), C--rich (filled squares) and extreme (circles) AGB sources plotted against their 8 \mic\ excesses. The lines show power-law fits for the dust mass-loss rate--excess relation for all three types of AGB candidates (solid line: O--rich, dashed line: C--rich, dot-dashed line: Extreme).
}
\label{vLMdotvsX8}
\end{figure}

\begin{figure}
\epsscale{0.8}\plotone{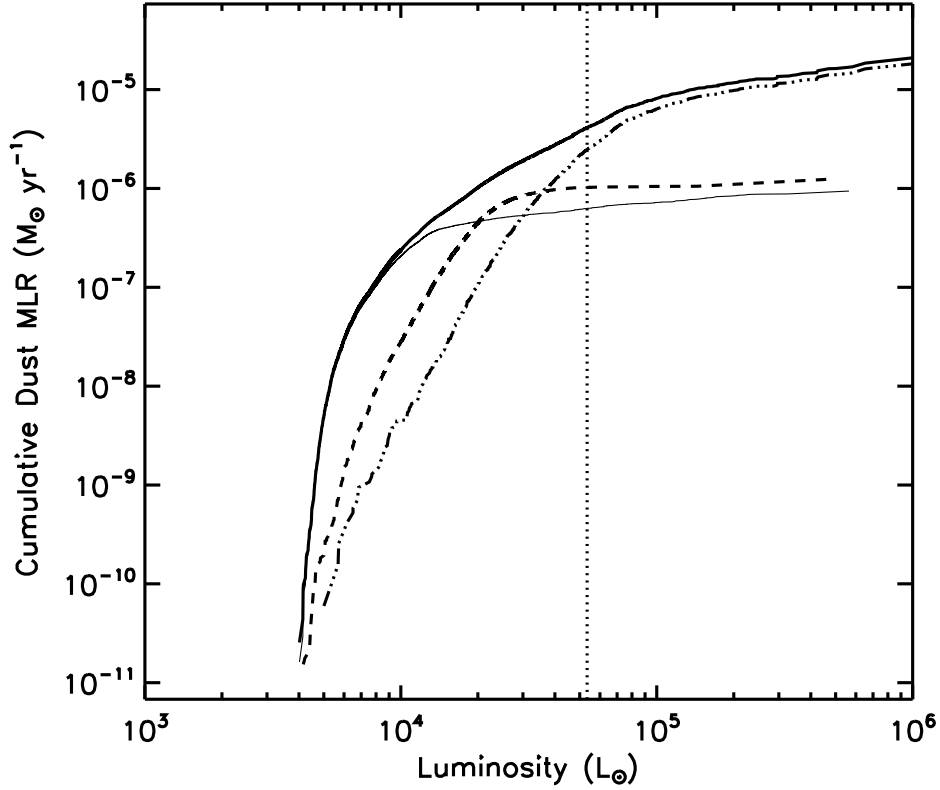}
\caption{The cumulative dust mass-loss rate as a function of luminosity for the O--rich (thin solid line), C--rich (dashed line) and Extreme (dot-dashed line) AGB stars in our lists. The thick solid line is the total AGB dust mass-loss rate as a function of luminosity. The extreme AGB stars are the major contributors to the mass-loss rate, at 2.36$\times 10^{-5}$ \msunperyr. The total dust mass-loss rate is 2.74$\times 10^{-5}$ \msunperyr.
The vertical dotted line is the classical AGB luminosity limit. As is evident from the figure, there are very few, if any, carbon--rich AGB candidates above this limit. However, deeply embedded extreme AGB stars and O--rich stars undergoing hot-bottom burning can exceed this luminosity limit.}
\label{mdotcontrib}
\end{figure}

We estimate the current LMC AGB star mass-loss budget using a method analogous to \citet{Blumetal07}. The MLRs obtained by \citet{vL1999} for spectroscopically identified AGB stars are plotted against their SAGE 8 \mic\ excess fluxes in Figure \ref{vLMdotvsX8}. The figure also shows power-law fits of the form
\ben
\nonumber
\log{\dot{M} (10^{-6} ~{\rm M}_\odot {\rm yr}^{-1})}=A+B\log{X_\nu({\rm mJy})} 
\een
to all three types of sources with the following best-fit parameters:
\ben
\nonumber
A=-1.5,\ B=1.0~\rm{(O-rich)}\\
\nonumber
A=-1.7,\ B=1.1~\rm{(C-rich)}\\
\nonumber
A=-2.8,\ B=1.7~\rm{(Extreme)}
\een
These relations are then used to derive MLRs for all the sources in our lists with reliable excesses. Figure \ref{mdotcontrib} shows the total dust MLR as a function of LMC AGB luminosity. The extreme AGB stars, despite their comparatively low numbers, are the most significant contributors to the AGB mass-loss budget in the LMC. We find the dust injection rates from O--rich, C--rich and extreme AGB stars to be 0.14, 0.24, and 2.36 $\times 10^{-5}$ \msunperyr\ respectively, which puts the estimate for the total dust injection rate from AGB stars at $2.74\times 10^{-5}$ \msunperyr. While stars with luminosities between $M_{bol}=-7.1$ and $M_{bol}\approx -7.8$ may be highly embedded AGB stars, sources brighter than $M_{bol}> -7.8$ are definitely supergiants \citep[see, {\it e.g.},][]{Sloanetal2009}. The luminosity distribution (Figure \ref{AGBLF}) shows that there are very few sources in this range, but it is clear from Figure \ref{mdotcontrib} that the contribution of the red supergiants to the dust injection rate is comparable to that of the AGB stars. This introduces a significant uncertainty in the value for the AGB dust injection rate derived in this section. 

Calculation of the total (gas+dust) injection rate requires knowledge of the gas:dust ratio, $\psi$, in the circumstellar shells of AGB stars. The lower metallicity of the LMC will result in a higher silicate gas:dust ratio but the carbon gas:dust ratio may be similar to Galactic values \citep{Habing1996}. Keeping this in mind, we choose gas:dust ratios of 200 and 500 for C--rich and O--rich stars respectively\footnote{The work of \citet{vL2008} suggests that $\psi$ scales in proportion to the metallicity for both O--rich and C--rich AGB stars, but this effect can not be accounted for in our simple estimate.}. The C--rich value is similar to that obtained by \citet{Skinneretal1999} for the Galactic carbon star, IRC+10 216. The gas MLR from the O--rich and C--rich AGB candidates is then 7$\times 10^{-4}$ \msunperyr\ and 4.8$\times 10^{-4}$ \msunperyr\ respectively. The chemical identification of the extreme AGB sources is not possible from our data, but assuming $\psi$=200 and $\psi$=500 for these stars will provide lower and upper limits. The gas MLR for the extreme AGB population is in the range 4.7--11.8$\times 10^{-3}$ \msunperyr. Thus, the total AGB gas MLR is (5.9--13)$\times 10^{-3}$ \msunperyr. We will improve on our estimate with our follow-up radiate transfer modeling of the dust shells around these stars using the 2DUST code \citep{UetaMeixner03}. The results of the present study will help constrain the range of optical depths and grain temperatures. The optical depth is an important input parameter for the code.

Taking into account the effects of variability on the excesses as discussed previously, we estimate that the MLR of an individual source can vary on average by a factor of $\sim$3 for the O--rich and  C--rich sources and $\sim$15 for the extreme AGB candidates. Due to the large number of O--rich and C--rich sources, the errors of the flux variations on the cumulative MLR estimates would probably cancel out. For the extreme AGB candidates, the larger effects of the variability and the smaller numbers may result in a significant change in the cumulative MLR. The magnitude of such an effect could be calibrated with an IR variability study of the LMC AGB stars to determine their periods and hence phase-correct the fluxes and excesses, but no such study exists for our complete sample.

Our C--rich AGB star gas MLR of 4.8$\times 10^{-4}$ \msunperyr\ is comparable to the value, 6$\times 10^{-4}$ \msunperyr, found by \citet{Matsuuraetal2009}. However, we note that \citet{Matsuuraetal2009} base their measurement on MLR versus color relations and include the extreme C--rich AGB stars, identified by IRS spectroscopy, as well as the color classified C--rich AGB stars. Our excess versus MLR relation method is two orders of magnitude more sensitive to lower MLR sources (dust MLRs of $\sim 10^{-11}$ \msunperyr) compared to the \citet{Matsuuraetal2009} method (dust MLRs of $\sim 10^{-9}$ \msunperyr). It is therefore fortuitous that our estimates are so similar and our exclusion of the highest MLR extreme C--rich AGB stars is balanced by the inclusion of more low MLR C--rich AGB stars. Furthermore, the total mass injection rate of C--rich mass loss is probably higher than both these estimates, but definitive numbers will require a clear identification of what fraction of the extreme AGB stars are C--rich.

Our estimate for the total LMC AGB gas MLR is higher than those obtained for two local group dwarf irregulars, WLM \citep[][(0.7--2.4)$\times 10^{-3}$ \msunperyr]{Jacksonetal200701} and IC 1613 \citep[][(0.2--1.0)$\times 10^{-3}$ \msunperyr]{Jacksonetal200702}. As their values for the MLRs of individual stars are in good agreement with the estimates of individual stars by \citet{vL1999} for the LMC, the disparity between our results and the values for WLM and IC 1613 could arise due to various reasons. The total MLR return due to AGB stars depends on the total number of AGB stars in a galaxy. The total stellar mass of the LMC \citep[$\sim 3 \times 10^9$ M$_\odot$,][]{vdM02} is two orders of magnitude greater than those estimated by \citet{Jacksonetal200701} and \citet{Jacksonetal200702} for WLM and IC 1613 ($1.1\times 10^7$ M$_\odot$ and $1.7\times 10^7$ M$_\odot$ respectively). Moreover, the age of the LMC bar is 4--6 Gyr \citep{Smecker-Haneetal2002}, which is optimal for an enhanced AGB population at present times, compared to the slightly older stellar population of WLM \citep[9 Gyr,][]{Jacksonetal200701}. Based on the differences in stellar masses and ages, we expect that the LMC has considerably more AGB stars than either WLM or IC 1316, leading to a higher total MLR from AGB stars. However, the differences in the MLR determination may also increase our MLR compared to theirs. Our IR excess method is able to detect sources with small MLR, while \citet{vL1999} and \citet{Jacksonetal200701} derive their estimates from the highest mass-losing sources alone. In this sense, ours is a more complete estimate of the total MLR. On the other hand, our interpolation of the \cite{vL1999} MLRs may not be appropriate for sources with very low excess. Thus, while the excess method is able to identify sources with low MLRs, the derived MLRs may be overestimates.

The current star formation rate (SFR) of the LMC is estimated to be about 0.1 \msunperyr\ \citep{sage3}, which is an order of magnitude higher than our calculated AGB MLR. Thus, the current star formation rate is not sustainable, assuming the LMC is a closed box system, and that AGB stars are the only means of replenishing the ISM. However, the most massive stars ({\it e.g.}, red supergiants, supernova explosions) may also contribute to the mass-loss return, and these need to be accounted for to complete the census of the mass budget return to the LMC. In addition, interactions with the Small Magellanic Cloud may cause infall or loss of mass that would need to be considered in the mass budget of the ISM.

\section{Summary and conclusions}
\label{fin}
\noindent We classify evolved stars in the SAGE Epoch 1 Archive based on their 2MASS and IRAC colors, and use photospheric models to estimate the infrared excess emission due to circumstellar dust. We obtain about 8200 O--rich, 5800 C--rich, and 1400 extreme AGB sources with reliable 8 \mic\ excesses (SNR $\geq$ 3). The corresponding numbers in the 24 \mic\ band are about 4700, 4900, and 1300 respectively. The excesses increase with an increase in luminosity in all four IRAC bands as well as the MIPS24 band. We use the 8 and 24 \mic\ excess fluxes to derive dust color temperatures and optical depths, and we observe that higher excess fluxes correspond to cooler temperatures and optically thicker dust shells.  These quantities for our list of AGB candidates with valid 8 \mic\ excesses are available in the form of electronic tables. We also estimate the present day AGB mass-loss budget in the LMC by comparing modeled mass-loss rates with our excesses estimates to find that the extreme AGB stars are the most significant contributors to mass loss in the LMC.  Our data also suggests that the rate of dust injection from red supergiants to the LMC ISM is comparable to the total dust injection rate from the AGB stars, which we calculate to be about $(5.9-13)\times 10^{-5}$ \msunperyr.

\begin{acknowledgements}
The authors wish to thank Rita Gautschy-Loidl and Peter Hauschildt for providing us with their photospheric models. We would also like to thank the anonymous referee for their helpful comments. The SAGE Project is supported by NASA/Spitzer grant 1275598 and NASA NAG5-12595.
\end{acknowledgements}

\clearpage 
\thispagestyle{empty}
\begin{landscape}
\begin{deluxetable}{rrrrrrrrrrrrrrrr}
\setlength{\tabcolsep}{0.05in}
\tablewidth{0pt}
\tablecolumns{16}
\tablecaption{Source List of Color-Selected AGB Star Candidates \label{SourceList}\tablenotemark{1}}
\tabletypesize{\scriptsize}
\tablehead{
\colhead{Identifier\tablenotemark{a}} &
\colhead{Type\tablenotemark{b}} &
\colhead{magU\tablenotemark{c}} & \colhead{$\delta$magU} & \colhead{magB} & \colhead{$\delta$magB} & \colhead{magV} & \colhead{$\delta$magV} & \colhead{magI} & \colhead{$\delta$magI} & \colhead{magJ} & \colhead{$\delta$magJ} & \colhead{magH} & \colhead{$\delta$magH} & \colhead{magK} & \colhead{$\delta$magK}
}
\startdata
SSTISAGE1A J054938.72--683458.2&O--rich&99.99&99.99&19.46&0.04&17.55&0.05&13.76&0.07&12.01&0.03&11.11&0.03&10.78&0.02\\
SSTISAGE1A J055530.35--684647.5&O--rich&20.46&0.19&18.11&0.03&16.18&0.03&13.95&0.04&12.59&0.02&11.72&0.02&11.48&0.02\\
SSTISAGE1A J055420.11--680449.5&O--rich&99.99&99.99&18.87&0.04&16.59&0.05&13.53&0.03&11.91&0.02&11.02&0.02&10.71&0.03\\
SSTISAGE1A J055729.20--684444.2&O--rich&17.97&0.06&17.47&0.04&16.32&0.08&16.38&0.30&11.93&0.02&11.03&0.02&10.75&0.02\\
SSTISAGE1A J055321.17--683114.7&O--rich&21.04&0.24&18.49&0.09&16.51&0.05&13.87&0.03&12.52&0.02&11.60&0.02&11.37&0.02\\
SSTISAGE1A J054522.57--684244.5&C--rich&99.99&99.99&20.68&0.06&16.45&0.03&13.82&0.04&12.49&0.02&11.17&0.02&10.38&0.02\\
SSTISAGE1A J055650.80--675030.5&C--rich&99.99&99.99&20.42&0.07&16.66&0.11&13.56&0.04&12.12&0.02&11.01&0.03&10.35&0.02\\
SSTISAGE1A J055835.31--682009.7&C--rich&99.99&99.99&20.13&0.05&16.28&0.03&13.55&0.04&12.30&0.02&11.02&0.03&10.20&0.02\\
SSTISAGE1A J055311.71--684720.9&C--rich&99.99&99.99&21.91&0.17&17.93&0.04&14.56&0.04&12.03&0.02&10.91&0.03&10.14&0.02\\
SSTISAGE1A J055036.67--682852.3&C--rich&21.61&0.44&19.31&0.05&16.68&0.03&14.42&0.04&12.87&0.03&11.92&0.03&11.52&0.02\\
SSTISAGE1A J052742.48--695251.5&Extreme&19.08&0.09&18.58&0.09&18.72&0.11&14.86&0.07&14.13&0.07&12.41&0.06&11.08&0.04\\
SSTISAGE1A J052714.19--695524.3&Extreme&20.04&0.12&19.65&0.07&18.36&0.05&14.45&0.05&13.74&0.03&12.13&0.03&10.85&0.02\\
SSTISAGE1A J053239.06--700157.5&Extreme&99.99&99.99&20.29&0.20&17.30&0.10&14.81&0.04&13.16&0.03&11.73&0.02&10.73&0.02\\
SSTISAGE1A J052950.52--700000.1&Extreme&18.80&0.09&18.28&0.09&17.40&0.14&16.08&0.05&13.96&0.04&12.30&0.04&10.80&0.03\\
SSTISAGE1A J053441.38--692630.7&Extreme&99.99&99.99&22.78&0.45&20.26&0.11&14.59&0.04&12.36&0.02&10.92&0.02&9.87&0.03\\
\enddata
\tablenotetext{1}{The electronic version of this table contains all the AGB candidates considered in this study.}
\tablenotetext{a}{SAGE Epoch 1 Archive (SAGE1A) designation, including position coordinates of IRAC source.}
\tablenotetext{b}{Sources are classified as O--rich or C--rich based on their J-\ks colors, or as ``Extreme" based on their 2MASS and IRAC colors. For more details, see \S \ref{data}.}
\tablenotetext{c}{Magnitudes and errors in the UBVI, JH\ks, IRAC and MIPS 24 \mic\ bands. A value of 99.99 in any band represents either a saturation or a non-detection.}
\end{deluxetable}
\clearpage

\thispagestyle{empty}
\begin{deluxetable}{rrrrrrrrrr}
\addtocounter{table}{-1}
\tablewidth{0pt}
\tablecolumns{10}
\tablecaption{\em (Continued)}
\tabletypesize{\scriptsize}
\tablehead{
\colhead{mag36} & \colhead{$\delta$mag36} & \colhead{mag45} & \colhead{$\delta$mag45} & \colhead{mag58} & \colhead{$\delta$mag58} & \colhead{mag80} & \colhead{$\delta$mag80} & \colhead{mag24} & \colhead{$\delta$mag24}\\
}
\startdata
10.53&0.03&10.72&0.03&10.50&0.04&10.44&0.05&9.86&0.07\\
11.34&0.03&11.40&0.03&11.28&0.04&11.19&0.04&10.59&0.14\\
10.47&0.04&10.64&0.02&10.46&0.05&10.40&0.04&9.97&0.07\\
10.47&0.03&10.60&0.02&10.39&0.03&10.30&0.04&10.01&0.08\\
11.07&0.07&11.02&0.04&10.90&0.03&10.76&0.03&10.10&0.08\\
9.67&0.03&9.70&0.03&9.68&0.04&9.30&0.04&8.89&0.04\\
9.76&0.04&9.86&0.03&9.81&0.04&9.30&0.04&9.26&0.03\\
9.34&0.05&9.03&0.02&8.80&0.03&8.68&0.03&8.61&0.03\\
9.51&0.05&9.43&0.04&9.27&0.04&8.99&0.03&8.66&0.03\\
11.06&0.04&11.08&0.04&10.89&0.04&10.68&0.05&10.21&0.09\\
9.02&0.05&8.65&0.04&8.39&0.03&8.00&0.03&7.49&0.02\\
9.62&0.04&9.12&0.03&8.71&0.03&8.24&0.03&7.61&0.03\\
9.38&0.04&9.33&0.03&9.19&0.04&8.83&0.03&8.46&0.03\\
8.94&0.04&8.23&0.04&7.55&0.03&6.85&0.02&6.03&0.02\\
9.06&0.05&8.33&0.05&7.81&0.02&7.29&0.03&6.74&0.02\\
\enddata
\end{deluxetable}

\clearpage

\thispagestyle{empty}
\begin{deluxetable}{rrrrrrrrrrrr}
\tablewidth{0pt}
\tablecolumns{12}
\tablecaption{Mid-infrared Excess Fluxes \label{ExcessList}\tablenotemark{1}}
\tabletypesize{\scriptsize}
\tablehead{
\colhead{Identifier} & \colhead{Type} & \colhead{X36\tablenotemark{a}} & \colhead{$\delta$X36\tablenotemark{b}} & \colhead{X45} & \colhead{$\delta$X45} & \colhead{X58} & \colhead{$\delta$X58} & \colhead{X80} & \colhead{$\delta$X80}& \colhead{X24}& \colhead{$\delta$X24}\\
&&\colhead{(mJy)}&\colhead{(mJy)}&\colhead{(mJy)}&\colhead{(mJy)}&\colhead{(mJy)}&\colhead{(mJy)}&\colhead{(mJy)}&\colhead{(mJy)}&\colhead{(mJy)}&\colhead{(mJy)}
}
\startdata
SSTISAGE1A J054938.72--683458.2&O--rich&4.546&0.622&1.918&0.315&2.338&0.287&1.335&0.210&0.457&0.051\\
SSTISAGE1A J055530.35--684647.5&O--rich&0.843&0.287&0.727&0.171&0.726&0.135&0.450&0.091&0.213&0.053\\
SSTISAGE1A J055420.11--680449.5&O--rich&4.254&0.784&1.932&0.279&2.170&0.363&1.222&0.187&0.349&0.050\\
SSTISAGE1A J055729.20--684444.2&O--rich&4.351&0.531&2.413&0.292&2.740&0.230&1.693&0.190&0.326&0.050\\
SSTISAGE1A J055321.17--683114.7&O--rich&2.387&0.654&2.392&0.268&1.909&0.163&1.313&0.100&0.428&0.049\\
SSTISAGE1A J054522.57--684244.5&C--rich&18.980&1.265&12.830&0.720&8.010&0.533&7.247&0.488&1.283&0.078\\
SSTISAGE1A J055650.80--675030.5&C--rich&13.170&1.375&7.974&0.645&5.065&0.509&6.534&0.504&0.606&0.048\\
SSTISAGE1A J055835.31--682009.7&C--rich&29.860&2.606&31.540&0.989&26.200&0.957&16.000&0.617&1.761&0.076\\
SSTISAGE1A J055311.71--684720.9&C--rich&19.900&1.948&16.800&1.066&13.170&0.819&9.984&0.475&1.570&0.077\\
SSTISAGE1A J055036.67--682852.3&C--rich&1.080&0.494&1.209&0.257&1.308&0.207&0.953&0.160&0.238&0.051\\
SSTISAGE1A J052742.48--695251.5&Extreme&69.590&2.958&62.220&2.082&50.750&1.422&40.600&1.094&7.258&0.160\\
SSTISAGE1A J052714.19--695524.3&Extreme&39.680&1.490&40.240&1.117&37.570&1.141&32.440&0.895&6.457&0.154\\
SSTISAGE1A J053239.06--700157.5&Extreme&49.920&1.958&33.430&0.812&24.320&0.834&18.790&0.459&2.950&0.068\\
SSTISAGE1A J052950.52--700000.1&Extreme&74.320&3.098&91.780&3.533&110.200&2.910&117.100&2.549&27.820&0.396\\
SSTISAGE1A J053441.38--692630.7&Extreme&66.460&3.197&83.610&3.607&86.330&1.828&77.850&2.089&14.490&0.305\\
\enddata
\tablenotetext{1}{The electronic version of this table contains all the AGB candidates with valid 8 \mic\ excesses.}
\tablenotetext{a}{The MIPS 24 \mic\ and IRAC 8.0, 5.8, 4.5, and 3.6 \mic\ band excess fluxes and errors in mJy. }
\tablenotetext{b}{The errors in the excess fluxes have been calculated by propagating the photometric errors:
\ben
\nonumber
\frac{\delta X_\nu}{X_\nu}=\sqrt{\left(\frac{\delta F_\nu}{F_\nu}\right)^2+\left(\frac{F^{mod}_\nu}{F^{mod}_{\rm H}}\right)^2\left(\frac{\delta F_{\rm H}}{F_{\rm H}}\right)^2}
\een
}
\end{deluxetable}
\end{landscape}

\clearpage

\begin{deluxetable}{rrrrrrrr}
\tablewidth{0pt}
\tablecolumns{8}
\tablecaption{Color temperatures and Optical Depths \label{TempTauList}\tablenotemark{1}}
\tabletypesize{\scriptsize}
\tablehead{
\colhead{Identifier}& \colhead{Type}& \colhead{T\tablenotemark{a}}& \colhead{$\delta$T}& \colhead{$ \tau$(8 \mic)\tablenotemark{b}}& \colhead{$\delta\tau$(8 \mic)}& \colhead{$\tau$(24 \mic)}& \colhead{$\delta\tau$(24 \mic)}\\
&&\colhead{(K)}&\colhead{(K)}&&&&
}
\startdata
SSTISAGE1A J054938.72--683458.2&O--rich&393&28&0.813&0.017&0.938&0.003\\
SSTISAGE1A J055530.35--684647.5&O--rich&352&37&0.806&0.028&0.935&0.005\\
SSTISAGE1A J055420.11--680449.5&O--rich&422&36&0.885&0.012&0.962&0.002\\
SSTISAGE1A J055729.20--684444.2&O--rich&502&48&0.916&0.008&0.972&0.002\\
SSTISAGE1A J055321.17--683114.7&O--rich&401&21&0.743&0.016&0.914&0.003\\
SSTISAGE1A J054522.57--684244.5&C--rich&525&26&0.922&0.004&0.974&0.001\\
SSTISAGE1A J055650.80--675030.5&C--rich&814&84&0.982&0.001&0.994&0.000\\
SSTISAGE1A J055835.31--682009.7&C--rich&705&32&0.953&0.002&0.985&0.000\\
SSTISAGE1A J055311.71--684720.9&C--rich&559&22&0.931&0.002&0.977&0.000\\
SSTISAGE1A J055036.67--682852.3&C--rich&446&53&0.949&0.007&0.983&0.001\\
SSTISAGE1A J052742.48--695251.5&Extreme&521&10&0.830&0.003&0.944&0.001\\
SSTISAGE1A J052714.19--695524.3&Extreme&495&9&0.793&0.004&0.931&0.001\\
SSTISAGE1A J053239.06--700157.5&Extreme&559&11&0.908&0.002&0.970&0.000\\
SSTISAGE1A J052950.52--700000.1&Extreme&456&5&0.601&0.005&0.867&0.001\\
SSTISAGE1A J053441.38--692630.7&Extreme&511&9&0.808&0.003&0.936&0.001\\
\enddata
\tablenotetext{1}{The electronic version of this table contains all the AGB candidates with valid 8 \mic\ excesses.}
\tablenotetext{a}{Color temperatures and related uncertainties, derived from the 8 \mic\ and 24 \mic\ excesses. The uncertainties are calculated by propagating the photometric errors.}
\tablenotetext{b}{Optical depths and related uncertainties at 24 \mic\ and 8 \mic\ derived from the color temperature. The uncertainties are calculated by propagating the photometric errors.}
\end{deluxetable}

\clearpage

\newpage

\newpage

\newpage

\newpage 

\newpage

\newpage

\newpage

\newpage

\newpage

\newpage

\newpage

\newpage

\newpage

\newpage

\newpage

\end{document}